\documentclass[fleqn,10pt]{wlscirep}
\usepackage[utf8]{inputenc}
\usepackage[T1]{fontenc}

\usepackage{longtable}

\newcommand\mnras{Mon. Not. R. Astron. Soc.}
\newcommand\apjl{Astrophys. J. Lett.}
\newcommand\apj{Astrophys. J.} 
\newcommand\aj{Astron. J.}                   % Astronomical Journal (the)
\newcommand\aap{Astron. Astrophys.}
\newcommand\apss{Astrophys. Space Sci.}    % Astrophysics and Space Science
\newcommand\araa{Annu. Rev. Astron. Astrophys.}   
\newcommand\pasp{Publ. Astron. Soc. Pac.}
\newcommand\apjs{Astrophys. J. Suppl.}               % Astrophysical Journal, Supplement

\title{An 18.9-minute Blue Large-Amplitude Pulsator Crossing the ``Hertzsprung Gap'' of Hot Subdwarfs}

\author[1]{Jie Lin}
\author[2,1]{Chengyuan Wu}
\author[1,3,*]{Xiaofeng Wang}
\author[4,5]{P\'eter N\'emeth}
\author[6,2]{Heran Xiong}
\author[2,7,8,9,10]{Tao Wu}
\author[11]{Alexei V. Filippenko}
\author[1]{Yongzhi Cai}
\author[11]{Thomas G. Brink}
\author[1]{Shengyu Yan}
\author[12]{Xiangyun Zeng}
\author[13]{Yangping Luo}
\author[1]{Danfeng Xiang}
\author[2]{Jujia Zhang}
\author[11]{WeiKang Zheng}
\author[11]{Yi Yang}
\author[1]{Jun Mo}
\author[1]{Gaobo Xi}
\author[14]{Jicheng Zhang}
\author[15]{Abdusamatjan Iskandar}
\author[15]{Ali Esamdin}
\author[16,9]{Xiaojun Jiang}
\author[1]{Hanna Sai}
\author[1]{Zixuan Wei}
\author[1]{Liyang Chen}
\author[1]{Fangzhou Guo}
\author[1]{Zhihao Chen}
\author[17]{Wenxiong Li}
\author[1]{Weili Lin}
\author[1]{Han Lin}
\author[1]{Xinghan Zhang}

\affil[1]{
Physics Department and Tsinghua Center for Astrophysics, Tsinghua University, Beijing, 100084, People's Republic of China
}
\affil[2]{
Yunnan Observatories, Chinese Academy of Sciences, Kunming, 650216, People's Republic of China
}
\affil[3]{
Beijing Planetarium, Beijing Academy of Sciences and Technology, Beijing, 100044, People's Republic of China
}
\affil[4]{
Astronomical Institute of the Czech Academy of Sciences, Fri\v{c}ova 298, Ond\v{r}ejov, 25165, Czech Republic
}
\affil[5]{
Astroserver.org, F\H{o} t\'er 1, Malomsok, 8533, Hungary
}
\affil[6]{
Research School of Astronomy \& Astrophysics, Australian National University, Canberra, 2611, Australia
}
\affil[7]{
Key Laboratory for the Structure and Evolution of Celestial Objects, Chinese Academy of Sciences, 396 Yangfangwang, Guandu District, Kunming, 650216, People's Republic of China
}
\affil[8]{
Center for Astronomical Mega-Science, Chinese Academy of Sciences, 20A Datun Road, Chaoyang District, Beijing, 100012, People's Republic of China
}
\affil[9]{
School of Astronomy and Space Science, University of Chinese Academy of Sciences, Beijing, 100049, People’s Republic of China
}
\affil[10]{
Institute of Theoretical Physics, Shanxi University, Taiyuan, 030006, China
}
\affil[11]{
Department of Astronomy, University of California, Berkeley, CA 94720-3411, USA 
}

\affil[12]{
Center for Astronomy and Space Sciences, China Three Gorges University, Yichang, 443000, People's Republic of China
}
\affil[13]{
Department of Astronomy, China West Normal University, Nanchong, 637002, People's Republic of China
}
\affil[14]{Department of Astronomy, Beijing Normal University, Beijing, 100875, People's Republic of China
}
\affil[15]{Xinjiang Astronomical Observatory, Chinese Academy of Sciences, Urumqi, 830011, People’s Republic of China
}
\affil[16]{
National Astronomical Observatories of China, Chinese Academy of Sciences, Beijing, 100012, People’s Republic of China
}
\affil[17]{
The School of Physics and Astronomy, Tel Aviv University, Tel Aviv, 69978, Israel
}

\affil[*]{wang\_xf@mail.tsinghua.edu.cn}

\begin{abstract}
Blue large-amplitude pulsators (BLAPs) represent a new and rare class of hot pulsating stars with unusually large amplitudes and short periods. 
Up to now, only 24 confirmed BLAPs \cite{Pietrukowicz+2017+OGLE_BLAPs,Kupfer+2019+high-g_BLAPs,Lin+etal+2021+tmts,McWhirter+Lam+2022+blap_candidates,Pigulski+Kolaczek-Szymanski+2022+TESS_BLAP,Ramsay+etal+2022+OW_blaps}  have been identified from more than one billion monitored stars, including a group with pulsation period longer than $\sim 20$\,min\cite{Pietrukowicz+2017+OGLE_BLAPs} (classical BLAPs, hereafter) and the other group with pulsation period below $\sim 8$\,min (high-gravity BLAPs\cite{Kupfer+2019+high-g_BLAPs}). 
The evolutionary path that could give rise to such kinds of stellar configurations is unclear.
Here we report on a comprehensive study of the peculiar BLAP discovered by the Tsinghua University -- Ma Huateng Telescopes for Survey (TMTS), TMTS~J035143.63+584504.2 (TMTS-BLAP-1) \cite{Lin+etal+2021+tmts}.
This new BLAP has an 18.9\,min pulsation period and is similar to the BLAPs with a low surface gravity and an extended helium-enriched envelope, suggesting that it is a classical BLAP \cite{Pietrukowicz+2017+OGLE_BLAPs}  at the shortest-period end.
In particular, the long-term monitoring data reveal that this pulsating star has an unusually large rate of period change, $\dot{P}/P=2.2 \times 10^{-6}\,{\rm yr^{-1}}$. 
Such a significant and positive value challenges its origins from both helium-core pre-white-dwarfs and core helium-burning subdwarfs, but is consistent with that derived from shell helium-burning subdwarfs. The particular pulsation period and unusual rate of period change indicate that TMTS-BLAP-1 is at a short-lived ($\sim 10^{6}$\,yr) phase of shell-helium ignition before the stable shell-helium burning \cite{Guo+2018+SHeB}; in other words, TMTS-BLAP-1 is going through a ``Hertzsprung gap'' of hot subdwarfs. The discovery of TMTS-BLAP1 possibly opens a new window to explore those rare pulsators that evolved from stripped-envelope stars.
\end{abstract}
\begin{document}

\flushbottom
\maketitle

\thispagestyle{empty}

\section{Main}
The minute-cadence observations by Tsinghua University -- Ma Huateng Telescopes for Survey (TMTS) enable the search of variable stars with periods shorter than 1\,hr. TMTS~J035143.63+584504.2 is such a newly discovered short-period variable, having an 18.9\,min period and a peak-to-peak amplitude  of $\sim 0.3$\,mag in white light \cite{Lin+etal+2021+tmts} (Fig.~\ref{fig:tmts_lightcurves}).
Follow-up observations with SNOVA, a 14\,inch telescope at Nanshan station of Xinjiang Observatories (Zhang J. et al., in prep.), confirmed its periodicity and revealed a clear sawtooth-shaped light curve (see Fig.~\ref{fig:folded_lc_sections}),  consistent with those of blue large-amplitude pulsators (BLAPs) discovered by the Optical Gravitational Lensing Experiment (OGLE) \cite{Pietrukowicz+2017+OGLE_BLAPs}. Moreover, the dereddened colour $(B_{\rm p}-R_{\rm p})_0=-0.47\pm0.19$\,mag and absolute magnitude $M_{\rm G}=1.43^{+0.18}_{-0.19}$\,mag derived from {\it Gaia} (E)DR3 data \cite{Gaia_Collaboration+2016+performance,Gaia_EDR3_content,Bailer-Jones+etal+2021+EDR3_distance,GaiaDR3+Apsis+parameters} support the classification of TMTS~J035143.63+584504.2 (hereafter TMTS-BLAP-1) as a BLAP \cite{Ramsay+2018+identify_gaia}.
This object is also bright in the ultraviolet bands \cite{Lanning+Meakes+1995+UV} and has recently been identified as a BLAP candidate (i.e., ZGP-BLAP-01) through the {\it Gaia} DR2 and Zwicky Transient Facility (ZTF) DR3 data \cite{McWhirter+Lam+2022+blap_candidates}. 

\begin{figure*}
\centering
    \includegraphics[width=0.6\textwidth]{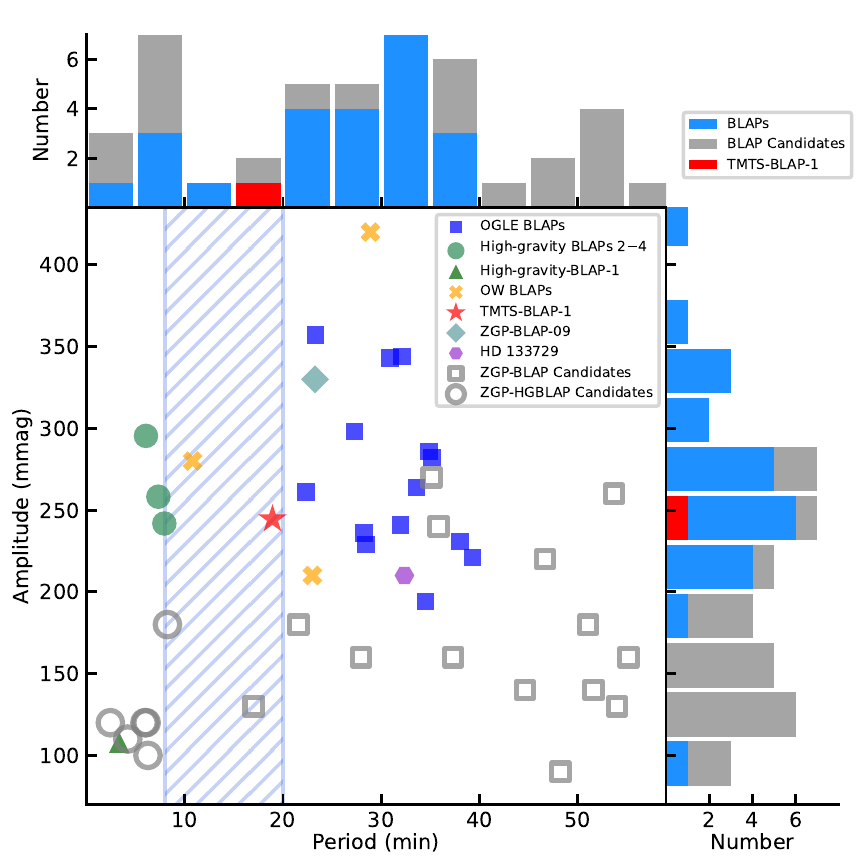}
    \caption{
    Distribution of all known BLAPs \cite{Pietrukowicz+2017+OGLE_BLAPs,Kupfer+2019+high-g_BLAPs,McWhirter+Lam+2022+blap_candidates,Pigulski+Kolaczek-Szymanski+2022+TESS_BLAP,Ramsay+etal+2022+OW_blaps} and candidates \cite{McWhirter+Lam+2022+blap_candidates} in the period$-$amplitude diagram. The OGLE BLAPs (blue squares) were measured in the OGLE $I$ band, the OW BLAPs (orange cross) were measured in the OW $g$ band, the amplitude of HD~133729 was derived from TESS band, and all other BLAPs and candidates were measured in the ZTF $r$ band. 
    The amplitude for HD~133729 is the intrinsic amplitude derived by correcting for the dilution from brighter companion \cite{Pigulski+Kolaczek-Szymanski+2022+TESS_BLAP}.
    For comparison, all amplitudes were converted into the peak-to-peak amplitude (rather than the semi-amplitude).
    The shaded area indicates the 8-to-20-minute gap\cite{Byrne+etal+2021+population}.
     }
    \label{fig:spectra_evolution}
\end{figure*}

The period--amplitude diagram of 24 confirmed BLAPs \cite{Pietrukowicz+2017+OGLE_BLAPs,Kupfer+2019+high-g_BLAPs,Lin+etal+2021+tmts,McWhirter+Lam+2022+blap_candidates,Pigulski+Kolaczek-Szymanski+2022+TESS_BLAP,Ramsay+etal+2022+OW_blaps} and 20 BLAP candidates \cite{McWhirter+Lam+2022+blap_candidates} is shown in Fig.~1.
The confirmed BLAPs tend to have higher pulsation amplitudes in comparison with those candidates.
There are several BLAPs located within the 8-to-20\,min period gap \cite{Byrne+etal+2021+population}, which can roughly divide the BLAP samples into classical (or low-gravity) and high-gravity groups. With the discovery of more BLAPs, the nominal period gap seems to become less distinct, while the scarcity in the 10-to-20-minute period is still noticeable (see the upper histogram of Fig.~1).
Since the boundary of period gap is arbitrary because of limited sample, TMTS-BLAP-1 likely represents a member of classical BLAPs at the shortest-period end, or a special BLAP between two groups, such as an intermediate-gravity BLAP \cite{Byrne+etal+2020+faint}.

\subsection{Phase-resolved spectroscopy}

\begin{figure*}
\centering
    \includegraphics[width=0.9\textwidth]{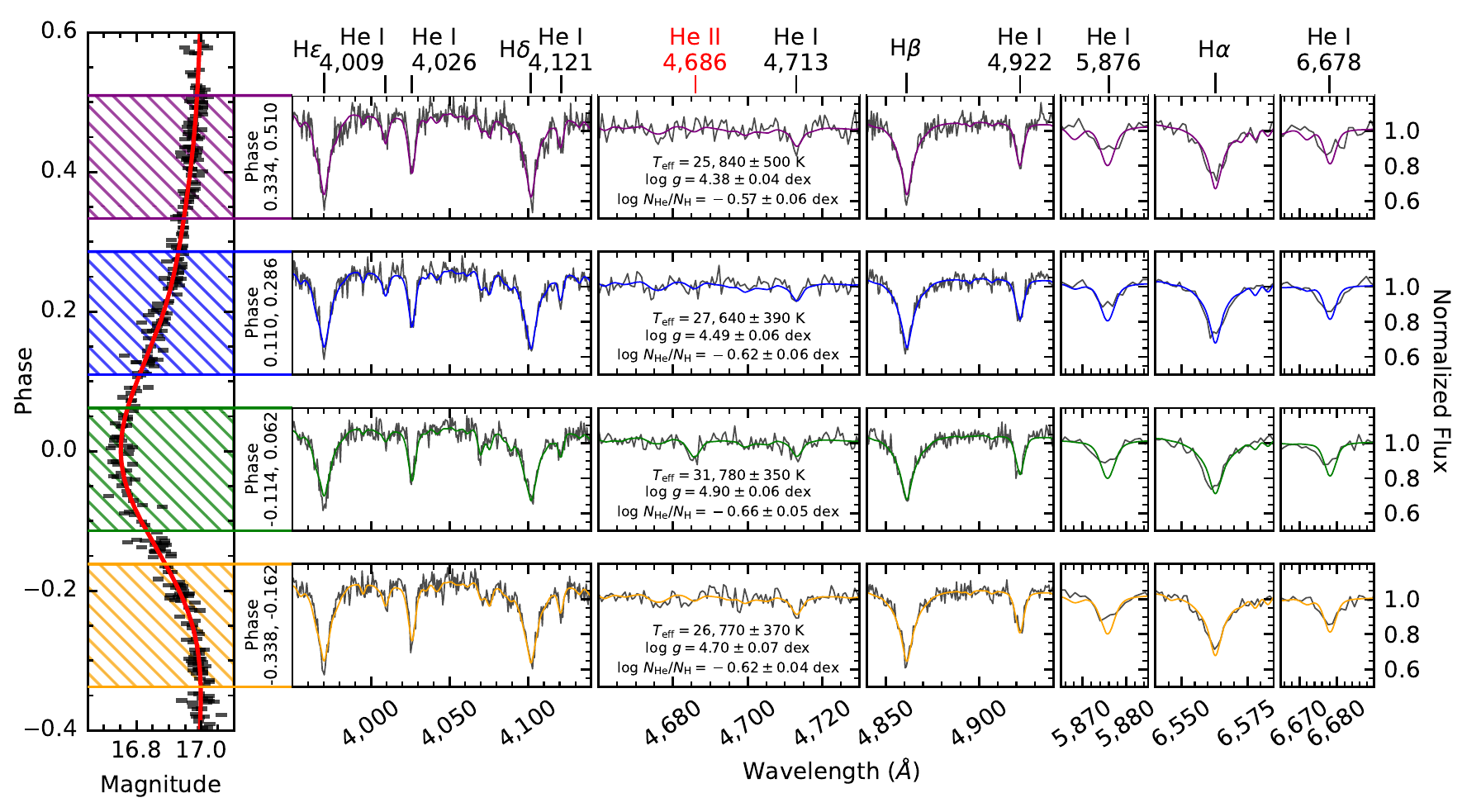}
    \caption{
    Keck LRIS spectra and best-fitting parameters for TMTS-BLAP-1 at four non-overlapping pulsation phases.
Left vertical panel shows the corrected ZTF $r$ folded light curve (black bars; see Method) with a best-fitting 3-harmonic Fourier model (red solid line). 
Four horizontal panels show the main spectral lines (black lines) with best-fitting stellar-atmosphere models (coloured lines). 
    \label{fig:spectra_evolution}
    }
\end{figure*}

To further solidify the classification of TMTS-BLAP-1, we obtained a series of phase-resolved spectra using the Low Resolution Imaging Spectrometer (LRIS) \cite{Oke+etal+1995+Keck+LRIS,McCarthy+etal+1998_LRIS} mounted on the 10\,m Keck~I telescope.
As shown in Fig.~2, the four spectra, with 200\,s exposure for each one, covered nearly a complete pulsation period.
The effective temperature of TMTS-BLAP-1 varied from 25,840\,K to 31,780\,K, with the highest temperature corresponding to the maximum brightness.
The change in radial velocity was found to precede the change in surface gravity by about a quarter of the pulsation period (see Fig.~\ref{fig:surface_pars}), supporting the idea that the TMTS-BLAP-1 pulsation is a radial mode \cite{Kupfer+2019+high-g_BLAPs}.
The surface gravity and moderately high helium abundance, inferred from the best-fitting atmospheric model, are consistent with those of the OGLE BLAPs \cite{Pietrukowicz+2017+OGLE_BLAPs}.
A noteworthy spectroscopic feature in TMTS-BLAP-1 is the emergence and vanishing of the He~II~$\lambda4686$ line during the 18.9\,min cycle, which seems to be also present in the OGLE BLAPs \cite{Pietrukowicz+2017+OGLE_BLAPs} but has not been observed in four high-gravity BLAPs with low helium abundance \cite{Kupfer+2019+high-g_BLAPs}.
While the changing visibility of the He~II~$\lambda4686$ line is a temperature effect, the observed difference in mean helium abundance between high-gravity and classical BLAPs may be regarded as an indicator of different evolutionary origins \cite{Zhang+Jeffery+2012+merge,Byrne+etal+2018+diffusion} or status.

\subsection{Rate of period change}

\begin{figure*}
\centering
    \includegraphics[width=0.6\textwidth]{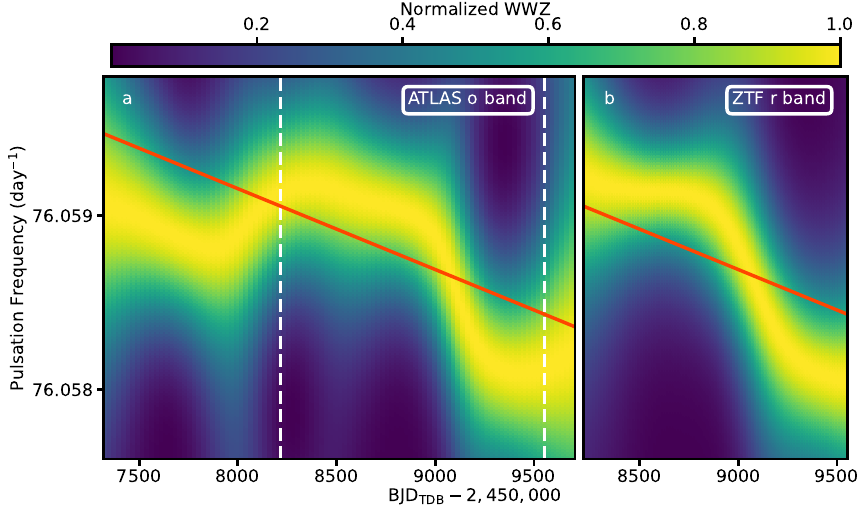}
    \caption{
Normalised weighted wavelet Z-transform (WWZ) plots for TMTS-BLAP-1.
The WWZ plots were computed from the ATLAS $o$ band (panel $a$) and the ZTF $r$ band (panel $b$). The WWZ powers across all frequencies in each time bin were normalised by the maximum power in the bin, and the frequency corresponding to the peak WWZ represents the pulsation frequency over a time interval. Each of the WWZ plots consists of 100 time bins and the frequency bin width is $2\times10^{-5}$\,day$^{-1}$.
The white dashed lines in panel $a$ indicate the time range corresponding to the observation coverage of the ZTF $r$ band.
The red solid lines denote the intrinsic pulsation frequency derived from the best-fitting LTTE model, while the fluctuation is believed to be caused by the Doppler effect of orbital motion.
    \label{fig:wwz}
    }
\end{figure*}

BLAPs are believed to stem from either helium-core pre-white-dwarfs (pre-WDs) or core helium-burning (CHeB) subdwarfs  \cite{Pietrukowicz+2017+OGLE_BLAPs, Romero+etal+2018+BLAPs, Byrne+etal+2018+BLAPs, Wu+etal+2018+CHeB,Kupfer+2019+high-g_BLAPs,Pigulski+Kolaczek-Szymanski+2022+TESS_BLAP}, since stars formed from these two channels cover the BLAP region in the Hertzsprung-Russell (HR) diagram.
The pre-WDs usually contract and cool while the CHeB stars burn helium steadily on nuclear timescales. 
To explore the nature of TMTS-BLAP-1, we tried to compute the precise rate of change of its pulsation period. Since the period vs. mean density relation (i.e., Ritter's relation \cite{Ritter+1879} ) is valid for BLAPs \cite{Romero+etal+2018+BLAPs,Kupfer+2019+high-g_BLAPs,Wu+etal+2018+CHeB}, it is used to trace the evolution of stellar radius \cite{Rodriguez-Segovia+etal+2022+cepheid_LMC}.

The variation of pulsation period of TMTS-BLAP-1 can be diagnosed using the weighted wavelet Z-transform (WWZ) \cite{Foster+1996+WWZ},
which is a practical technique for visualising time-dependent periodicity in observation data.
As shown in Fig.~3, the long-term observations from both the Asteroid Terrestrial-impact Last Alert System (ATLAS) \cite{Tonry+etal+2018+ATLAS,Heinze+etal+2018+ATLAS_variables} and ZTF \cite{ZTF+2019+first,ZTF+2019+products} surveys support a consistent pulsation-frequency variation in TMTS-BLAP-1.
The pulsation frequency exhibits a large-amplitude variation superimposed on a declining trend.
The complicated variability of pulsation period means that the rate of period change of TMTS-BLAP-1 cannot be reliably determined by the simple method as used in Pietrukowicz et al. (2017) \cite{Pietrukowicz+2017+OGLE_BLAPs}, where the rates were calculated directly based on the difference of pulsation periods between two different epochs.
The overall frequency trend in the WWZ plot suggests an average rate of period change as $\dot{P}/P \approx 2.1\times 10^{-6}\,{\rm yr}^{-1}$ in about 6.5 yrs. 

\begin{figure*}
\centering
    \includegraphics[width=0.6\textwidth]{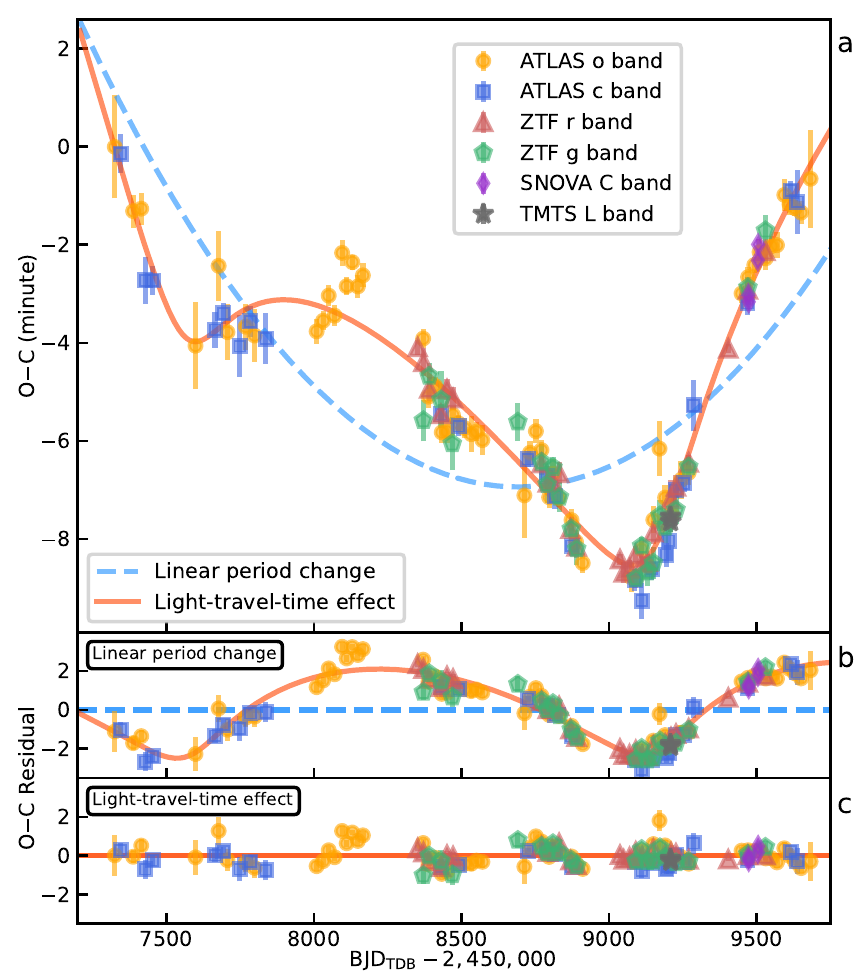}
    \caption{
    $O-C$ diagram for the pulsation period of TMTS-BLAP-1.
    The observed times of maximum light ($T_{\rm max}^O$) were obtained from the subsets of ATLAS, ZTF, SNOVA, and TMTS observations. 
    Each subset of the ATLAS and ZTF data covers up to 20\,days, while each subset of the TMTS and SNOVA data corresponds to an individual night.
    The $O-C$ variability is modeled by assuming a linear period change (blue dashed line) and LTTE caused by orbital motion (red solid line). Their residuals are shown in panels $b$ and $c$, respectively. Error bars represent 1$\sigma$ confidence throughout this paper.
    \label{fig:O-C}
    }
\end{figure*}

In order to determine precisely the rate of change of the pulsation period from the pulsation-frequency fluctuation, we plotted the $O-C$ diagram for TMTS-BLAP-1 using data from ZTF, ATLAS, SNOVA, and TMTS (see Method). $O-C$ is a strong diagnostic tool for evaluating the discord 
between times of a given event (e.g., peak of pulsation) and predicted values from a stable and accurate clock \cite{Sterken+2005+O-C}.
We computed the values of $O-C=T_{\rm max}^O-T_{\rm max}^C$ following the ephemeris
\begin{equation}
T_{\rm max}^C(E)={\rm BJD_{TDB}}2,457,325.0469+0.0131477151\times E ,
\label{Eq:ephemeris}
\end{equation}
where $E$ is the cycle number elapsed from the initial epoch, and $T_{\rm max}^O$ and $T_{\rm max}^C$ represent the observed and calculated times of maximum light, respectively.
A complete $O-C$ diagram is shown in Fig.~4, where the overall trend supports that TMTS-BLAP-1 has a high and positive rate of period change.
The $O-C$ diagram agrees with the trend and superimposed variation seen in the WWZ plot.
To examine the reliability of our analysis, we also computed the $O-C$ diagram for another newly confirmed BLAP, the 23.3\,min ZGP-BLAP-09 (see Fig.~\ref{fig:O-C_ztfblap}), and we did not find any significant period change in its $O-C$ diagram.

Assuming that the variations in the $O-C$ diagram of TMTS-BLAP-1 is caused only by a linear pulsation period change due to stellar evolution (see Method), we obtained a rate of period change equal to $\dot{P}/P = 2.19\pm 0.19 \times 10^{-6}\,{\rm yr}^{-1}$  from the best-fitting $O-C$ model.
A cyclic feature in the $O-C$ residuals is shown in the panel $b$ of Fig.~4. Similar cyclic behaviours have been revealed recently in another BLAP, HD~133729 \cite{Pigulski+Kolaczek-Szymanski+2022+TESS_BLAP}.
Since stellar evolution theory favours the origin of BLAPs in a binary system \cite{Pietrukowicz+2017+OGLE_BLAPs,Pigulski+Kolaczek-Szymanski+2022+TESS_BLAP}, the cyclic feature seen in the $O-C$ diagram of TMTS-BLAP-1 is very likely caused by the light-travel-time effect (LTTE) induced from the orbital motion \cite{Irwin+1952+LiTE,Pigulski+Kolaczek-Szymanski+2022+TESS_BLAP}. 
By assuming that the pulsating star orbits the barycentre of the binary system, the $O-C$ variability was fitted with the light-travel-time model of elliptical orbits \cite{Irwin+1952+LiTE} (see Method). The best-fitting model can be used to correct the phase-folded light curves and obtain coherent light-curve shapes (see Fig.~\ref{fig:folded_lc}). The model suggests an orbital period of $P_{\rm orb}=1576\pm18$\,days and a mass function of
$f(M_2)=(1.22\pm0.12)\times10^{-3}\,{\rm M}_\odot$ (see details in Table~\ref{tab:params_omc}). 
The long orbital period marginally matches the orbital period distribution of hot subdwarfs derived from stable Roche lobe overflow (RLOF) channel \cite{Han+etal+2002+sdB}.
The derived mass function suggests that TMTS-BLAP-1 could be orbited by a low-mass star or a brown dwarf,
which is below the mass limit allowed by the classical critical mass ratio (i.e., $q\sim 1.5$). However, this critical mass ratio is uncertain and it could be much higher (e.g., $q_{\rm th}\sim$ 20.0) due to the thermal equilibrium mass loss \cite{Ge+etal+2020_qth}, which allows the formation of hot subdwarfs with wide-orbit low-mass companions.
Otherwise, a possibly low orbital inclination can also be responsible for the low value of the mass function \cite{Han+etal+2002+sdB, Morales-Rueda+etal+2003+sdb_period,Kupfer+etal+2015+sdB_period}.
The LTTE model cannot explain all details in the $O-C$ diagram; the excess could be caused by fluctuations from a potential third star/planet.
The best-fitting model also provides a more reliable estimate of the rate of period change, $\dot{P}/P = 2.23\pm 0.09 \times 10^{-6}\,{\rm yr}^{-1}$, consistent with the estimate from the WWZ analysis.
Owing to the overall trend of $O-C$ variability, no matter what model is selected to fit the $O-C$ diagram, a positive and high rate of period change is inevitable.
At such a large rate, the evolutionary timescale of TMTS-BLAP-1 will be only $P/\dot{P}\approx4.5\times10^5$\,yr. 
If some BLAPs evolve rapidly from 10 to 20 minutes, this rapid evolution within the ``period gap'' can also help explain the observed scarcity of BLAPs with 10-to-20-minute pulsation periods, in addition to the possible contribution from mass gap of pre-WDs \cite{Byrne+etal+2021+population}.

\begin{table}[h]
\caption{Best-fitting parameters for $O-C$ diagram of TMTS-BLAP-1. \label{tab:params_omc}} 
\centering
\begin{tabular}{lcc}
\hline
Model   &  Linear period change   & LTTE   \\
\hline
Rate of period change, $\dot{P}/{P}$ ($\rm 10^{-6}\,yr^{-1}$)& $2.19\pm0.19$ & $2.23\pm0.09$ \\
Orbital period, $P_{\rm orb}$ (days)& & $1576.32\pm17.85$ \\
Projected semimajor axis, $a_1$\,sin\,$i$ (au)& & $0.283\pm0.009$ \\
Eccentricity, $e$ & & $0.526\pm0.050$ \\
Time of periastron passage, $T_0$ & & BJD$_{\rm TDB}$ $2,459,164.1\pm26.9$ \\
Longitude of periastron, $\omega$ & & $298\pm9^o$ \\
Mass function, $f(M)\,(10^{-3}\,{\rm M}_\odot)$ & & $1.22\pm0.12 $ \\
Offset for initial epoch, $\Delta T_0$ (min) & $ 1.11\pm0.59$   &$-0.18\pm0.28$ \\
Offset for pulsation period, $\Delta P_0$ (s)    &$-0.00930\pm0.00076$ & $-0.00935\pm0.00033$ \\
Systematic uncertainty, $\sigma_f$ (min)    &$1.7\pm0.1$ &$0.50\pm0.03$ \\
\hline
\end{tabular}
\end{table}

\subsection{Discussion}
The significant positive rate of period change indicates that TMTS-BLAP-1 evolves toward larger stars, inconsistent with the trend predicted by cooling process of pre-WDs ($\dot{P}/P \lesssim -10^{-7}\,{\rm yr^{-1}}$ ).
On the other hand, the rates derived from core helium-burning (CHeB) stars are initially very small ($\lvert \dot{P}/P \rvert \lesssim 10^{-8}\,{\rm yr^{-1}}$ ) while their convective helium cores are growing slowly. But, at the late time of the central helium-burning phase, the stars begin to shrink owing to the deficit of generated heating energy, resulting in accelerated decay of their pulsation periods.

After exhaustion of the central helium in a CHeB star, its CO core begins to shrink and the central temperature decreases as a result of neutrino energy loss, whereas its helium envelope expands simultaneously because the energy from core contraction cannot be radiated rapidly from the envelope. During this stage, the shell-helium burning (SHeB) is unstable and the evolution of its stellar structure is on the Kelvin-Helmholtz timescale, analogous with post-main-sequence stars in the Hertzsprung gap (e.g., subgiants). Hence, this stage can be defined as the ``Hertzsprung gap" of hot subdwarfs. 
The hot subdwarfs usually represent the core or shell helium-burning stars around the blue end of the horizontal branch \cite{Heber+2009+araa}, while the Hertzsprung gap represents the short-lived phase of post-main-sequence stars before they climb on the giant branch and start stable shell-hydrogen burning. We adopted the term ``Hertzsprung gap'' here to denote an analogous rapidly-evolving stage of hot subdwarfs before stable shell-helium burning. The``Hertzsprung gap'' does not mean that there is a gap seen in the HR diagram for the hot subdwarfs, because of various masses of hydrogen-enriched envelope.

With persistent increase in the central density, the CO core of the star finally evolves into a degenerate state, and then the contraction of the core stops. Thereafter, stable helium burning develops on the surface of the CO core; the star enters into a stable shell helium-burning phase, analogous with that of red giant branch (RGB) stars.
The SHeB is a neglected scenario \cite{Kupfer+2019+high-g_BLAPs} proposed to understand the physical origin of (at least a part of) classical BLAPs, which predicts that the rate of change of the pulsation period will be very large for some potential BLAPs if they are crossing the ``Hertzsprung gap''. More detailed SHeB models developed from binary evolution channel for BLAPs will be presented by Xiong H. et al. (in prep.).

\begin{figure*}
\centering
    \includegraphics[width=0.9\textwidth]{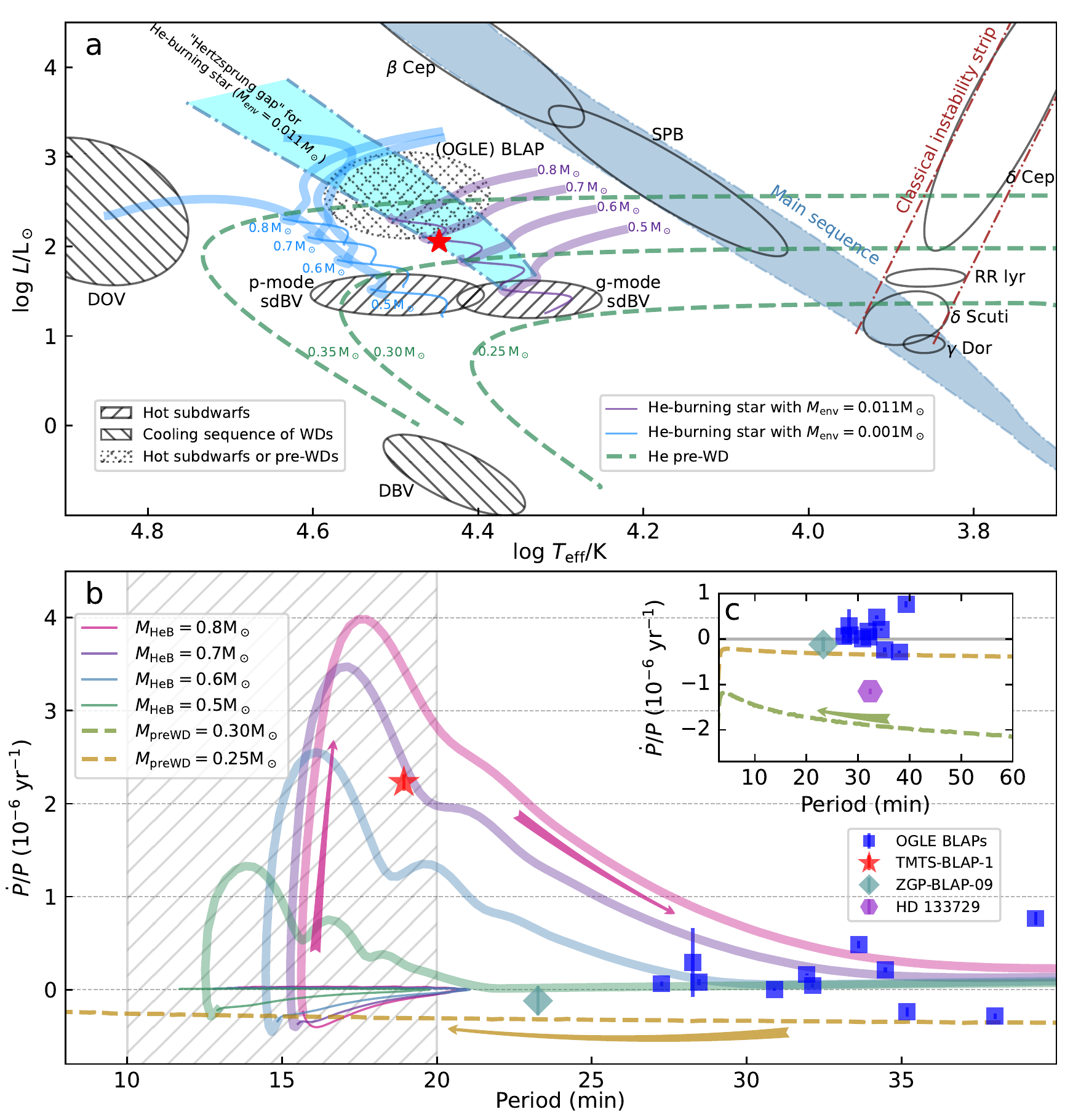}
    \caption{
    Hertzsprung-Russell diagram and period vs. rate of period change diagram for three candidate BLAP models.
    (a) Helium-core pre-WD (dashed lines), CHeB (thin solid lines), and SHeB (thick solid lines) models are overplotted on the HR diagram.
    The CHeB and SHeB models represent two different stages of helium-burning stars.
    The regions of pulsating stars refer to Paxton et al. (2019) \cite{Paxton+etal+2019+MESA}, in which the BLAP region is designed based on the OGLE BLAPs \cite{Pietrukowicz+2017+OGLE_BLAPs}. 
    Beneath the main sequence, the cooling sequence and helium-burning stars are shaded by different patterns.
    (b, c) Pre-WD (dashed lines), CHeB (thin solid lines), and SHeB (thick solid lines) are shown in the $P$--$\dot{P}/P$ diagram.
    The hydrogen envelope masses for helium-burning stars were assumed to be  $0.011\,{\rm M}_\odot$.
    The rates for OGLE BLAPs were taken from Pietrukowicz et al. (2017) \cite{Pietrukowicz+2017+OGLE_BLAPs}.
    They were roughly estimated by the difference of pulsation periods between two different epochs, and should be refined by $O-C$ diagrams in future research if one wants to further investigate their physical origins. The rate of HD~133729 was provided by Pigulski et al. (2022) \cite{Pigulski+Kolaczek-Szymanski+2022+TESS_BLAP}. 
    The shaded area denotes a rough period range where the BLAPs are significantly less populated. The arrows indicate the direction of evolution.
    \label{fig:HR_PPdot}
    }
\end{figure*}

The evolutionary tracks from all three candidate models (helium-core pre-WD, CHeB, and SHeB) are presented in Fig.~5, where one can see that the CHeB and SHeB models correspond to two distinguishable phases of helium-burning stars. All three models can explain the observed luminosities and effective temperatures of BLAPs,  and TMTS-BLAP-1 is located exactly in the ``Hertzsprung gap'' of helium-burning stars with a hydrogen envelope mass $M_{\rm env}=0.011\,{\rm M}_\odot$.
The helium-core pre-WD model predicts that the mass of TMTS-BLAP-1 is 0.30--0.35$\,{\rm M}_\odot$, while the CHeB and SHeB models predict much larger values, i.e.  $\sim 0.7\,{\rm M}_\odot$. The different masses inferred from the above three models provide an approach of constraining the physical origin of BLAPs \cite{Kupfer+2019+high-g_BLAPs}, but the stellar masses of BLAPs cannot be measured precisely from current observations because of their strong dependencies on the highly uncertain measurements of surface gravity.

Since the discovery of BLAPs \cite{Pietrukowicz+2017+OGLE_BLAPs}, ``BLAP'' has been used as a convenient term for all pulsating stars with high effective temperature and relatively large pulsation amplitude. 
Obviously, it is difficult to conclude that all BLAPs share the same physical origin, given that they have different observational features including pulsation period, surface gravity, absolute magnitude, helium abundance, and light-curve shape.
Since BLAPs locate in a common region in the HR diagram for different candidate models, it is necessary to explore their origins using the diagnostic tools other than the HR diagram.

The $P$--$\dot{P}$ diagram is a very efficient tool probing the physical origins of pulsars \cite{Lorimer+Kramer+2012+pulsar}. Here we adopted a similar approach, the $P$--$\dot{P}/P$ diagram, to expose the differences among three candidate models.
As shown in the lower panel of Fig.~5, a $P$--$\dot{P}/P$ diagram was plotted for a dozen of known BLAPs along with the models.
It seems difficult to model the distribution of BLAPs in the diagram by single evolutionary origin.
An outstanding feature of SHeB is that the star can expand rapidly during the  short-lived phase of shell-helium ignition \cite{Guo+2018+SHeB} (i.e., ``Hertzsprung gap'' of hot subdwarfs), 
which thus leads to a prominent rate of period change for pulsation period $\lesssim 20$\,min if their hydrogen envelope masses are similar to those of typical sdB stars.
 The rate of period change of TMTS-BLAP-1 favours its origin of an about 0.7\,${\rm M}_\odot$ SHeB star with an envelope mass $M_{\rm env}=0.011\,{\rm M}_\odot$, consistent with the results derived from the HR diagram. Whereas neither CHeB nor helium-core pre-WD can explain such a large and positive rate of period change.
 However, the mass estimated from the helium-burning models may have a large uncertainty, because the mass of the hydrogen envelope cannot be accurately determined.
If some BLAPs are SHeB stars, the rapid evolution of SHeB stars crossing the ``Hertzsprung gap'' will help explain the observed scarcity of BLAPs with pulsation periods below 20\,min.
 Alternatively, surface-gravity/period gap between two groups of BLAPs have been also interpreted as a result of fewer helium-core pre-WDs in the intermediate mass range \cite{Byrne+etal+2021+population}, with the assumption that BLAPs come from helium-core pre-WDs. 

It is not unexpected that some BLAPs correspond to shell helium-burning subdwarf stars. An outstanding feature in the HR diagram is that stars evolved from the hydrogen main sequence have produced plentiful pulsator categories, but the pulsating stars corresponding to the descendants of ``helium main-sequence stars'' (here including core helium-burning stars with very thin hydrogen envelopes) have hardly been revealed in the past.
The evolution of hot subdwarfs could be analogous with those of well-known hydrogen main-sequence stars, since the main difference between core helium-burning and core hydrogen-burning stars is the dominant element of fuel that determines the type of nuclear fusion driven in their cores. Meanwhile, some hot subdwarfs can also climb on the giant branch and finally cool down as white dwarfs.

Current existing pulsator categories are all based on observational facts of new pulsating variable stars, implying that we may miss potential classes from rare pulsating stars,
especially those corresponding to short-lived stages of stripped-envelope stars. TMTS-BLAP-1 provides an interesting observational evidence that some hot subdwarfs can leave their core helium-burning stage and appear as pulsating variables in a distinct region of the HR diagram. With the Legacy Survey of Space and Time (LSST) and other wide-field facilities in the future, more previously unknown pulsating stars evolved from stripped-envelope stars will be discovered, and they will further improve our understanding of pulsating variable stars and hot subdwarfs.

\section{Methods}
\subsection{Photometric observations from TMTS and Gaia}
The minute-cadence observations from TMTS can reveal periodic variability as short as 10\,min for objects brighter than 18\,mag \cite{Zhang+etal+2020+tmts,Lin+etal+2021+tmts}.
TMTS-BLAP-1 was captured by TMTS when they monitored two adjacent plates from the Large Sky Area Multi-Object Fiber Spectroscopy Telescope (LAMOST) on December 24 and 25, 2020 \cite{Lin+etal+2021+tmts}.
The ultrashort pulsation period of TMTS-BLAP-1 was revealed by a Lomb–Scargle periodogram \cite{Lomb+1976,Scargle+1982,VanderPlas+2018+LSP_understanding} using {\sc python} package {\sc gatspy}\cite{VanderPlas+etal+2015+gatspy,Vanderplas+2015+zenodo} and the pulsation amplitude was automatically obtained from the best-fitting Fourier model through the \emph{TMTS Lightcurve Analysis Pipeline} \cite{Lin+etal+2021+tmts} (see Fig.~\ref{fig:tmts_lightcurves}).

\emph{TMTS Lightcurve Analysis Pipeline} can estimate a preliminary distance, extinction, dereddened colour, and absolute magnitude for targets based on the {\it Gaia} DR2 data \cite{Gaia_Collaboration+2016+performance,Gaia_collaboration+2018+data}.
However, these estimates are rough.
Hence, we applied more reliable methods to obtain these values.
Owing to the low signal-to-noise ratio of the parallax of TMTS-BLAP-1 obtained from the {\it Gaia} EDR3 database, we adopted the photogeometric distance, $4.37^{+0.69}_{-0.61}$\,kpc, from the catalog of Bailer-Jones et al. (2021) \cite{Bailer-Jones+etal+2021+EDR3_distance}.
The photogeometric distances were calculated using parallax, color and apparent magnitude given in {\it Gaia} EDR3 catalog, as these parameters can give more accurate distance estimates for stars with poor parallaxes \cite{Bailer-Jones+etal+2021+EDR3_distance}.
The {\it Gaia} DR3 provides stellar parameters for more than 470\,million sources \cite{GaiaDR3+Apsis+parameters}, so the interstellar extinction for TMTS-BLAP-1 can be estimated based on measured $E({\rm B_P-R_P})$ and $A_{\rm G}$ values from sources in the same sky area.
We extracted a subset within $0.5^\circ$ of TMTS-BLAP-1.
The sources in the subset without reliable parallax measurements (i.e., 
$\varpi/\sigma_\varpi > 5.0$) were excluded.
Because $E({\rm B_P-R_P})$ depends on the intrinsic colour of the source, we further selected sources with $(B_{\rm p}-R_{\rm p})_0 < 0.5$ mag \cite{Ramsay+2018+identify_gaia}. 
From the 20 nearest sources around TMTS-BLAP-1 (selected by  $1/\varpi$), we obtained  $E{\rm (B_{\rm p}-R_{\rm p})} = 1.24\pm 0.19$\,mag and $A_{\rm G} = 2.28\pm 0.35$\,mag for TMTS-BLAP-1.
Thus, its dereddened colour is $(B_{\rm p}-R_{\rm p})_0=-0.47\pm0.19$\,mag, and its absolute magnitude is $M_{\rm G}=1.43^{+0.18}_{-0.19}$\,mag.

\subsection{Spectra and Spectral Energy Distribution} 
To avoid a low signal-to-noise ratio (SNR) and also minimise the Doppler smearing due to the fast spectral evolution of BLAPs, phase-resolved spectroscopy for these pulsating stars must be obtained from high-time-resolution observations on large-aperture telescopes. 
For this reason, only several BLAPs \cite{Pietrukowicz+2017+OGLE_BLAPs,Kupfer+2019+high-g_BLAPs,McWhirter+Lam+2022+blap_candidates} had phase-resolved spectroscopic observations before this work.

We obtained a series of four spectra of TMTS-BLAP-1 using the 10\,m Keck~I telescope and the Low Resolution Imaging Spectrometer (LRIS) \cite{Oke+etal+1995+Keck+LRIS,McCarthy+etal+1998_LRIS} instrument (blue grism 600/4000 with resolving power $R=940$; red grating 400/8500 with resolving power $R=930$). 
The spectra were observed at four different pulsation phases on 8 September 2021 (UT), with each having an exposure time of 200\,s and a readout time of $\sim 50$\,s. 
The spectra were reduced by the dedicated pipeline  {\sl LPipe} \cite{Perley+2019+lris}, following standard procedures: corrections for bias and flat field, removal of cosmic rays, extraction of the one-dimensional spectrum, wavelength calibration through comparison lamps, and flux calibration through observations of spectrophotometric standard stars. 
The strongest telluric absorption bands were also removed from the spectra using the standard-star spectra.
The spectra show strong Ca absorption lines, which are likely of interstellar origin. Our analysis confirms that the Ca K line does not move with the atmosphere.

The atmospheric parameters were determined by fitting non-local thermodynamic equilibrium {\sc Tlusty} (v207) and {\sc Synspec} (v53) \cite{hubeny17, lanz07} models to each individual spectrum. 
The iterative spectral analysis procedure ({\sc XTgrid}; \cite{nemeth12}) applies a steepest-descent $\chi^2$ minimisation to simultaneously optimise all free parameters and search for the best-fitting model. 
The models included H, He, C, N, O, Mg, Si, and Fe opacities in the atmospheric-structure calculations as well as in the spectrum synthesis.
{\sc XTgrid} calculates new models on the fly and adjusts the model parameters and atomic data input to link the variations in the theoretical atmospheric structure to the observable emergent spectrum precisely.
All comparisons were done globally using the entire observed spectral range and a piecewise normalisation of the model to the observation.
During this iterative search, the effective temperature,
surface gravity, chemical abundance, and projected rotational velocity are adjusted independently for minimising the global $\chi^2$.
In parallel, the radial velocity was also determined by shifting each observation to the model. 
The procedure converges once the relative changes of all model parameters and $\chi^2$ drop below 0.5\% in three consecutive iterations.
Next, parameter errors were measured by mapping the parameter space around the solution, including correlations between effective temperature and surface gravity. 
A fit to the blue part of a single LRIS observation is shown in Fig.~\ref{fig:nlte_fitting}.
The atmospheric parameters from the four LRIS spectra are listed in Table~\ref{tab:params}, and the (periodic) variations in these surface parameters are shown in Fig.~\ref{fig:surface_pars}.

Even though {\sc XTgrid} is able to fit the projected rotation velocity, it is usually difficult to be measured from low-resolution spectra. 
In addition, we encountered a serious degeneracy among rotation, surface gravity, and spectral resolution. 
The better-than-average seeing ($\sim 0.7''$) resulted in a higher resolution, exceeding the nominal resolution of the LRIS. 
Therefore, for consistency, we decided to keep the projected rotation velocity at 0. 
Likewise, we neglected microturbulence; its effects are unmeasurable in our spectra.

Fig.~\ref{fig:SED_fitting} shows the spectral energy distribution (SED) of TMTS-BLAP-1 from the Lyman limit to 50,000\,\AA.
{\sc Tlusty} provides the SED from the same models used to calculate the spectrum synthesis. 
The theoretical SED was convolved with various filter passbands and compared to observed data from broad-band photometry. 
For this comparison, photometric observations were collected within $2''$ around TMTS-BLAP-1 from the VizieR Photometry Viewer service and listed in Table.~\ref{tab:sed}.
The interstellar extinction toward TMTS-BLAP-1 is high, $E(B-V) = 0.87$\,mag according to the dust maps of Schlafly et al. (2011) \cite{schlafly11}. 
The SED fit suggests a lower value, $E(B-V) \approx 0.70$\,mag.
Using the distance derived from the {\it Gaia} EDR3 data \cite{Bailer-Jones+etal+2021+EDR3_distance}, we estimated that the average luminosity of TMTS-BLAP-1 is $113^{+15}_{-12}\,{\rm L}_\odot$.

\subsection{Weighted wavelet Z-transform analysis and $O-C$ diagram} 

There are more than 6\,yr observations from the ATLAS Forced Photometry Server \cite{Tonry+etal+2018+ATLAS,Heinze+etal+2018+ATLAS_variables} and the ZTF Public Data Release 10 (DR10) \cite{ZTF+2019+first,ZTF+2019+products}, which allows a best constraint on the rate of period change of TMTS-BLAP-1. In order to obtain accurate photometric measurements from ATLAS observations, the observation epochs with SNR $<10.0$ or reduced $\chi^2 > 2.0$ were excluded. For ZTF data, the detections with $catflag=32768$ were all excluded.
The Modified Julian Days (MJDs) of data are all converted into Barycentric Julian Dates with Barycentric Dynamical Time ($\rm BJD_{TDB}$).

We had tried to calculate the rate of period change for TMTS-BLAP-1 using the simple method introduced by Pietrukowicz et al. (2017) \cite{Pietrukowicz+2017+OGLE_BLAPs}. In their method, the rates of period change were directly estimated by the difference of pulsation periods between two different epochs. However, this results in multiple different values of $\dot{P}/P$ (from about $10^{-6}\,{\rm yr^{-1}}$ to $ 8\times10^{-6}\,{\rm yr^{-1}}$) for TMTS-BLAP-1, dependent on how the observational data are divided into two segments.
These various $\dot{P}/P$ values have reliable SNR and are supported from both the ZTF and ATLAS data, and thus cannot be simply explained by statistical or systematic errors. Therefore, we realised that the observed rate of period change of TMTS-BLAP-1 may not be a constant, and the rate of period change due to the stellar evolution is modulated by some extra effects. So, we adopted two stronger techniques, WWZ analysis and the $O-C$ diagram, to diagnose and reveal the rate of period change for TMTS-BLAP-1.

We performed the WWZ analysis for the ATLAS $o$-band and the ZTF $r$-band light curves using the {\sc python} package {\sc libwwz}.
The time-scale for determining the decay constant is set to 600\,days in the procedure.
The number of time bins for each WWZ plot is set to 100 and the frequency bin width is $2\times10^{-5}$\,cycle\,day$^{-1}$. 
The WWZ powers across all frequencies in each time bin were divided by the maximum power in the bin.

The $O-C$ diagram is difficult to compute when the observation cadence is much longer than the photometric period of the target. 
In order to bring out the details of period changes for TMTS-BLAP-1, we developed a method to estimate the times of maximum light ($T_{\rm max}^O$) from ATLAS and ZTF survey observations.
First, we fitted the entire light curve in each band using the 3-harmonic Fourier model,
\begin{equation}
{\rm Mag}\,(A_k,\phi_k, t)= A_0+\sum\limits_{k=1}^3  A_k \, \sin (2\pi  \,k\,f\, t + \phi_k) ,
\end{equation}
where $t$ represents the time of observation and $f$ is the pulsation frequency inferred from the Lomb-Scargle periodogram; $A_k$ and $\phi_k$ represent the Fourier amplitudes and phases, respectively. These Fourier parameters uniquely determine the light-curve shapes in each band.
Then the ATLAS and ZTF data were split into 20\,day subsets, and the subsets with fewer than 5 data points were excluded. This results in 132 subsets.
However, owing to the deficit of data points in individual subsets, the light curve from each subset is difficult to be fitted with a model that includes 7 free parameters.

By assuming that the light-curve shapes have not changed significantly over the survey observations, the light curves of subsets can be modeled using fixed Fourier parameters ($A_k$ and $\phi_k$) with a free parameter $\Phi$ for phase offset, namely ${\rm Mag'}\,(\Phi, t ) \rvert_{ t}=  {\rm Mag}\,( A_k, \phi_k, t+\Phi/f ) \rvert_{A_k,\phi_k, t} $. The hypothesis about the light-curve shapes can be verified below by the coherent profiles of folded light curves after correction (see Fig.~\ref{fig:folded_lc}).
 We fitted the light curves from each subset using the model ${\rm Mag'}\,(\Phi, t )$ and obtained the times of maximum light ($T_{\rm max}^O$) from the best-fitting models. The $T_{\rm max}^O$ closest to the mid-time of each subset were taken as the times of maximum light for computing $O-C$. 
 
 With $O-C=T_{\rm max}^O-T_{\rm max}^C$, we obtained a preliminary $O-C$ diagram using the $T_{\rm max}^O$ above.
The Fourier parameters and light-curve shapes were determined by assuming a constant pulsation period. Owing to the period changes over the survey observations, the amplitudes of light curves were actually underestimated (see panels $a,c$ in Fig.~\ref{fig:folded_lc}). In order to obtain more-accurate light-curve shapes in each band, we modeled the preliminary $O-C$ diagram and obtained a new ephemeris to correct the pulsation phase of each epoch. This results in much more coherent pulsation profiles for phase-folded light curves (see panels $b,d$ in Fig.~\ref{fig:folded_lc}). 
Hence, the peak-to-peak pulsation amplitudes for the ZTF $r$ band and the ATLAS $o$ band are $245\pm3$\,mmag and $245\pm4$\,mmag, respectively.
The Fourier parameters derived from corrected phase-folded light curves were applied to compute the times of maximum light ($T_{\rm max}^O$) again. Finally, we obtained the $O-C$ diagram in Fig.~4.
The minute-cadence observations from SNOVA and TMTS also contributed several $T_{\rm max}^O$, which were obtained by fitting the subsets corresponding to individual nights.
All times of maximum light are shown in Table.~\ref{tab:Tmax}. The phase-folded light curves and best-fitting models for every subset are shown in Fig.~\ref{fig:folded_lc_sections}.

In the case of a linear period change, the $O-C$ model can be given as a function of cycle number $E$ \cite{Sterken+2005+O-C},
\begin{equation}
(O-C)_{\rm linear}= \Delta T_0 + \Delta P_0 E + \frac{1}{2} \frac{\dot{P}}{P} \Bar{P}^2E^2 ,  
\label{Eq:O-C_linear}
\end{equation}
where $\Delta T_0$ and $\Delta P_0$ are the offsets for initial epoch and pulsation period in the ephemeris (i.e., Eq.~\ref{Eq:ephemeris}), respectively. $\Bar{P}$ is the average pulsation period over the whole time interval, which is inferred directly from the Lomb–Scargle periodogram. 
In the case of LTTE induced by orbital motion, the $O-C$ can be expressed as \cite{Irwin+1952+LiTE,Huseyin+etal+2021+twocb,Song+etal+2019+twocs_sdb,Hajdu+etal+2022+triple}
\begin{equation}
(O-C)_{\rm LTTE}=(O-C)_{\rm linear}+ a_1\,\sin{i}\frac{1-e^2}{1+e\,\cos{\nu}}\,\sin{(\nu+\omega)} ,
\label{Eq:O-C_ltte}
\end{equation}
where $(O-C)_{\rm linear}$ represents the contribution from pulsation period change due to stellar evolution (Eq.~\ref{Eq:O-C_linear}), $a_1\,\sin{i}$ is the projected semimajor axis of the absolute orbit, $e$ and $\omega$ are the eccentricity and longitude of periastron (respectively), and $\nu$ is the true anomaly (which is the function of eccentricity $e$, orbital period $P_{\rm orb}$, time of periastron passage $T_0$, and time of observations $t$ \cite{Irwin+1952+LiTE}).
In order to fit the $O-C$ diagram using the models in Eq.~\ref{Eq:O-C_linear} and Eq.~\ref{Eq:O-C_ltte}, we introduced a free parameter $\sigma_f$ for offsetting the systematic uncertainties and  adopted the likelihood function $\mathcal{L}$ introduced by Go{\'z}dziewski et al. (2015) \cite{Gozdziewski+etal+2015+likelihood}, namely
\begin{equation}
\log \mathcal{L}=-\frac{1}{2}  \sum_i^N \frac{(O-C)_{{\rm res},i}^2}{\sigma_i^2+\sigma_f^2}
- \sum_i^N \log \sqrt{{\sigma_i^2+\sigma_f^2}} -N \log{\sqrt{2\pi}} ,
\label{Eq:likelihood}
\end{equation}
where $(O-C)_{{\rm res},i}$ denotes the residual of $i$th calculated $(O-C)$ minus the model (i.e., Eq.~\ref{Eq:O-C_linear} and Eq.~\ref{Eq:O-C_ltte}).
$\sigma_i$ represents the error of $i$th calculated $(O-C)$ and $N$ is the total number of $(O-C)$.
The best-fitting parameters from both models are listed in Table~\ref{tab:params_omc}.
With the best-fitting LTTE model and ephemeris given by Eq.~\ref{Eq:ephemeris}, the pulsation period of TMTS-BLAP-1 is 
$18.9325540\pm0.0000055$\,min on ${\rm BJD_{TDB}}\,2,457,325$ and increases with a rate of $\dot{P}=4.22 \pm 0.17 \times 10^{-5}\,{\rm min\,yr^{-1}}$ .

For comparison, we also computed the $O-C$ diagram for another new BLAP, ZGP-BLAP-09 (Fig.~\ref{fig:O-C_ztfblap}).
Its $O-C$ diagram is almost constant and presents obviously less variability than that of TMTS-BLAP-1, implying that the rate of period change of ZGP-BLAP-09 is small and is not modulated by the similar cyclic behaviour in TMTS-BLAP-1. 
That allows us fit the $O-C$ values using the model inferred from linear period change (i.e., Eq.~\ref{Eq:O-C_linear}). The best-fitting model derives an insignificant rate of period change, $\dot{P}/P = -1.2\pm 1.6 \times 10^{-7}\,{\rm yr}^{-1}$, for ZGP-BLAP-09.

\subsection{Stellar Evolution Models}

In order to explore the unusual observational properties of BLAPs, we have performed calculations for all three candidate physical models using the state-of-art stellar evolutionary code Modules for Experiments in Stellar Astrophysics ({\sc MESA}; version: 12115) \cite{Paxton+etal+2011+MESA,Paxton+etal+2013+MESA,Paxton+etal+2015+MESA,Paxton+etal+2018+MESA,Paxton+etal+2019+MESA}. According to the HR diagram and $T_{\rm eff}$--$\log\,g$ diagram from BLAPs \cite{Kupfer+2019+high-g_BLAPs}, these pulsators between main-sequence and sdB stars may stem from the core-contraction phase of helium WDs with massive cores (e.g., $M_{\rm core} > 0.25\,{\rm M}_\odot$) or from low-mass helium-burning stars ($M \lesssim 1\,{\rm M}_\odot$).
In our calculations, we adopted the OP opacity tables \cite{Seaton+2005+OPgs98}. However, atomic diffusion and turbulent mixing can change the abundance for subdwarf B stars \cite{Michaud+etal+2011+sdB, Hu+etal+2011+diffusion}, while the rotational mixing and gravitational settling compete with each other to affect the atmospheric composition of pre-ELM WDs \cite{Istrate+etal+2016+ELM}. Both atomic diffusion and radiative levitation can lead to opacity-driven pulsations in post-common envelope objects with effective temperatures similar to BLAPs \cite{Byrne+etal+2018+BLAPs,Byrne+etal+2018+diffusion}. Nevertheless, we do not attempt to include the atomic diffusion, radiative levitation, and additional mixing mechanism in our current models; detailed calculations are very complicated and time-consuming.

To construct the helium-core pre-WD models, we evolved $\rm 1.8\,{\rm M}_\odot$ solar-metallicity ($Z=0.02$) main-sequence stars up to the red giant branch until their helium cores reached the mass thresholds (e.g., $0.25$ and $0.35\,{\rm M}_\odot$) in a mass step of $\Delta{\rm M}=0.01\,{\rm M}_\odot$. The hydrogen envelope is artificially removed by a extremely rapid mass-loss rate ($2.0\times{10}^{-4}\,{\rm M}_\odot\,{\rm yr}^{-1}$) \cite{Xiong+etal+2017,Wu+etal+2018+CHeB} until the envelope masses are less than $0.01\,{\rm M}_\odot$. Afterward, we replace the rapid mass-loss rate with a classical Reimers' wind \cite{Reimers+1975+wind} during the core-contraction phase until their luminosities decrease below $1\,{\rm L}_\odot$.

In order to obtain the helium-burning models, including CHeB and (unstable and stable) SHeB phases, and avoid the uncertainties in binary evolution, 
we constructed solar-metallicity ($Z=0.02$) zero-age CHeB stars with mass in the range of 0.5--2.0\,${\rm M}_\odot$. 
The hydrogen envelope with mass of 0.001--0.011\,${\rm M}_\odot$ is added onto the surface of the naked helium core by accretion \cite{Pelisoli+etal+2021+NA}. 
All nuclear reactions inside stars are shut off during the accretion phase and restored afterward. 
The stellar evolution terminates at $\log\,g = 3.8$ owing to the extreme time consumption during the helium giant phase, or terminates on the WD cooling track.
The CHeB phase corresponds to the stage when the star has a central helium-burning convective core, where SHeB stars have a contracting CO core (unstable helium-shell burning) or a degenerate CO core (stable helium-shell burning).
As introduced above, whether the central helium is exhausted is the criterion for differentiating the CHeB and SHeB phases, whereas the unstable and stable SHeB phases are distinguished by whether the helium burning zone moves outward.
The boundaries between successive phases are clearly shown as blue dashed-dotted lines in Fig.~4. The cyan area between lines corresponds to the ``Hertzsprung gap" for the helium-burning stars.

\subsection{Pulsation Period}

BLAPs are thought to be excited by $\kappa$-mechanism due to the Z-bump (also named iron-group elements) \cite{Pietrukowicz+2017+OGLE_BLAPs,Kupfer+2019+high-g_BLAPs,Romero+etal+2018+BLAPs,Byrne+etal+2020+faint},
which is enhanced through the action of radiative levitation \cite{Charpinet+etal+1997+driving,Fontaine+etal+2003+driving,Byrne+etal+2018+BLAPs,Byrne+etal+2020+faint}.
We suggest that some BLAPs, whether they are from CHeB or SHeB channels, have similar excitation mechanisms.

The variations seen in $\log\,g$ and RV of TMTS-BLAP-1 (Fig.~\ref{fig:surface_pars}) support the previous assumption that pulsations of BLAPs are in the radial fundamental mode \cite{Pietrukowicz+2017+OGLE_BLAPs,Kupfer+2019+high-g_BLAPs,Wu+etal+2018+CHeB},
we thus adopt the Ritter’s relation \cite{Ritter+1879} to estimate the pulsation period.
This relation yielded for pulsating stars connects pulsation period with stellar mean density. Hence, the pulsation period $P$ of BLAPs can be computed through available stellar parameters from evolutionary tracks \cite{Kupfer+2019+high-g_BLAPs}, namely 
\begin{equation}
P=\frac{2\pi}{f\omega_{\rm dyn}}= \frac{2\pi}{f}\, \left(\frac{G\,M}{R^3}\right)^{-\frac{1}{2}} ,
\end{equation}
where $\omega_{\rm dyn}$, $M$, and $R$ are the stellar dynamical frequency, mass, and radius, respectively. Since the oscillations in BLAPs are the fundamental mode, we adopted the dimensionless frequency $f=3.725$ (i.e., the median of $f\approx$3.65--3.8 \cite{Kupfer+2019+high-g_BLAPs}).
The rates of period change were calculated using $r=\Delta P / \Delta t$, where $\Delta P$ and $\Delta t$ represent the period and age difference between successive nodes on evolutionary tracks, respectively. To test the validity of Ritter’s relation and that the pulsation of radial fundamental mode can be efficiently excited, we further perform the  asteroseismic analysis for the helium-burning model with $M_{\rm core}=0.7\,{\rm M}_\odot$ and $M_{\rm env}=0.01\,{\rm M}_\odot$ by adopting the oscillation code  GYRE \cite{Townsend+Teitler+2013+GYRE,Townsend+etal+2018+GYRE,Goldstein+Townsend+2020+pulsation_GYRE}.  For radial fundamental modes of both adiabatic and non-adiabatic pulsations, we find that the Ritter’s relation is well consistent with the model calculations within an accuracy of 8\%. The asteroseismic analysis indicates that the radial fundamental modes can be excited for both CHeB and SHeB phases even if the atomic diffusion and radiative levitation processes are not included. 

\section{Data availability}
The ZTF $r$- and $g$-band photometry can be obtained from the NASA/IPAC Infrared Science Archive (\url{https://irsa.ipac.caltech.edu}).
The ATLAS $o$- and $c$-band magnitudes can be obtained from the ATLAS forced photometry server (\url{https://fallingstar-data.com/forcedphot}). 
All reduced light curves and spectra used for this work are available at our Zenodo page ( \url{https://doi.org/10.5281/zenodo.6425425}) .
Some evolutionary tracks can also be obtained from the Zenodo page. 

\section{Code availability}
The codes of {\sc Tlusty} (v207) and {\sc Synspec} (v53) that are used for generating (non-LTE) model atmospheres and producing synthetic spectra are available at \url{https://www.as.arizona.edu/~hubeny}, and the services of online spectral analyses ({\sc XTgrid}) are provided from Astroserver (\url{www.Astroserver.org}). The {\sc python} package {\sc libwwz} (v1.2.0) for WWZ analysis can be obtained from \url{https://pypi.org/project/libwwz}. The general tools for timing analysis are provided from {\sc python} package {\sc gatspy} (v0.3) (\url{http://www.astroml.org/gatspy} or \url{https://zenodo.org/record/47887}).
The software {\sc MESA} (v12115) used for stellar evolutionary calculations is available at \url{http://mesastar.org}.

\section{Acknowledgments}
We are grateful to Pawel Pietrukowicz for sharing very useful information about OGLE BLAPs.
The work of X.-F.W. is supported by the National Science Foundation of China (NSFC grants 12033003 and 11633002), the Ma Huateng Foundation, the Scholar Program of Beijing Academy of Science and Technology (DZ:BS202002), and the Tencent Xplorer Prize. 
C.-Y.W. is supported by the National Natural Science Foundation of China (NSFC grant 12003013).
P.N. acknowledges support from the Grant Agency of the Czech Republic (GA\v{C}R 22-34467S) and from the Polish National Science Centre under projects UMO-2017/26/E/ST9/00703 and UMO-2017/25/B/ST9/02218.
The Astronomical Institute in Ond\v{r}ejov is supported by the project RVO:67985815.
T.W. is grateful for support from the B-type Strategic Priority Program of the Chinese Academy of Sciences (grant XDB41000000), from the National Key R\&D Program of China (grant 2021YFA1600402), from the NSFC of China (grants 11873084 and 12133011), from the Youth Innovation Promotion Association of the Chinese Academy of Sciences, and from the Ten Thousand Talents Program of Yunnan for Top-notch Young Talents. T.W. also acknowledges with gratitude the computing time granted by the Yunnan Observatories, and provided on the facilities at the Yunnan Observatories Supercomputing Platform and the ``PHOENIX Supercomputing Platform'' jointly operated by the Binary Population Synthesis Group and The Stellar Astrophysics Group at Yunnan Observatories, Chinese Academy of Sciences.
A.V.F.'s group at U.C. Berkeley received financial support from the Miller Institute for Basic Research in Science (where A.V.F. was a Miller Senior Fellow), the Christopher R. Redlich Fund, and many individual donors.
Y.-Z.C. is funded by the China Postdoctoral Science Foundation (grant 2021M691821). 

Some of the data presented herein were obtained at the W. M. Keck Observatory, which is operated as a scientific partnership among the California Institute of Technology, the University of California, and NASA; the observatory was made possible by the generous financial support of the W. M. Keck Foundation.

This work has made use of data from the Asteroid Terrestrial-impact Last Alert System (ATLAS) project. The ATLAS project is primarily funded to search for near-Earth objects (NEOs) through NASA grants NN12AR55G, 80NSSC18K0284, and 80NSSC18K1575; byproducts of the NEO search include images and catalogs from the survey area. This work was partially funded by Kepler/K2 grant J1944/80NSSC19K0112 and {\it HST} GO-15889, and STFC grants ST/T000198/1 and ST/S006109/1. The ATLAS science products have been made possible through the contributions of the University of Hawaii Institute for Astronomy, the Queen’s University Belfast, the Space Telescope Science Institute, the South African Astronomical Observatory, and The Millennium Institute of Astrophysics (MAS), Chile.

Based in part on observations obtained with the Samuel Oschin 48\,inch telescope and the 60\,inch telescope at Palomar Observatory as part of the Zwicky Transient Facility project. ZTF is supported by the U.S. National Science Foundation under grant AST-1440341, AST-2034437  and a collaboration including Caltech, IPAC, the Weizmann Institute for Science, the Oskar Klein Center at Stockholm University, the University of Maryland, the University of Washington, Deutsches Elektronen-Synchrotron and Humboldt University, Los Alamos National Laboratories, the TANGO Consortium of Taiwan, the University of Wisconsin at Milwaukee, Trinity College Dublin, Lawrence Livermore National Laboratories, Lawrence Berkeley National Laboratory and IN2P3, France. Operations are conducted by COO, IPAC, and UW.

This research has used the services of \mbox{\url{www.Astroserver.org}} under reference UMVVTX. 
It has also used the VizieR catalogue access tool, CDS, Strasbourg, France (DOI: 10.26093/cds/vizier).
We also used data from the European Space Agency (ESA) mission
{\it Gaia} (\url{https://www.cosmos.esa.int/gaia}), processed by the {\it Gaia}
Data Processing and Analysis Consortium (DPAC,
\url{https://www.cosmos.esa.int/web/gaia/dpac/consortium}). Funding for the DPAC
has been provided by national institutions, in particular the institutions
participating in the {\it Gaia} Multilateral Agreement.
This work used the International Variable Star Index (VSX)\cite{Watson+etal+2006+VSX} database, operated at AAVSO, Cambridge, Massachusetts, USA.

\section*{Author Contributions}
J.L., C.-Y.W., X.-F.W., and P.N. drafted the manuscript; A.V.F., T.W., and Y.-Z.C. also helped with the manuscript, and A.V.F. edited it in detail. X.-F.W. is the PI of TMTS and SNOVA.
J. L. discovered this source by analysing the large-volume data from TMTS observations and performed the timing analysis to determine its rate of period change.
C.-Y.W. computed the stellar evolution models for helium-burning stars and helium-core pre-WDs, and H.-R.X. provided some key ideas for these models. 
T.W. contributed to the asteroseismic theory and analysis.
P.N. determined the atmospheric parameters from Keck-I LRIS spectra; Y.-Z.C., S.-Y.Y., Y.-P.L., and D.-F.X. assisted in the spectral analysis.
The Keck-I LRIS spectra were provided by A.V.F.’s group (including A.V.F, T.G.B., W.-K.Z., and Y.Y.). 
A.I., A.E., and J.-J.Z. contributed to the observations with SNOVA and the Lijiang 2.4\,m telescope, and X.-Y.Z. reduced these data.
J.M., G.-B.X., J.-C.Z., and J.L. contributed to the building, pipeline, and database of TMTS.
G.-B.X., J.M., X.-J.J., H.-N.S., Z.-X.W., L.-Y.C., F.-Z.G., Z.-H.C., W.-X.L., W.-L.L., H.L., and X.-H.Z. contributed to the operations of TMTS.

\section*{Competing interests}
The authors declare no competing interests.

\appendix
\section{Extended Data}

\begin{figure*}[h]
\centering
    \includegraphics[width=0.98\textwidth]{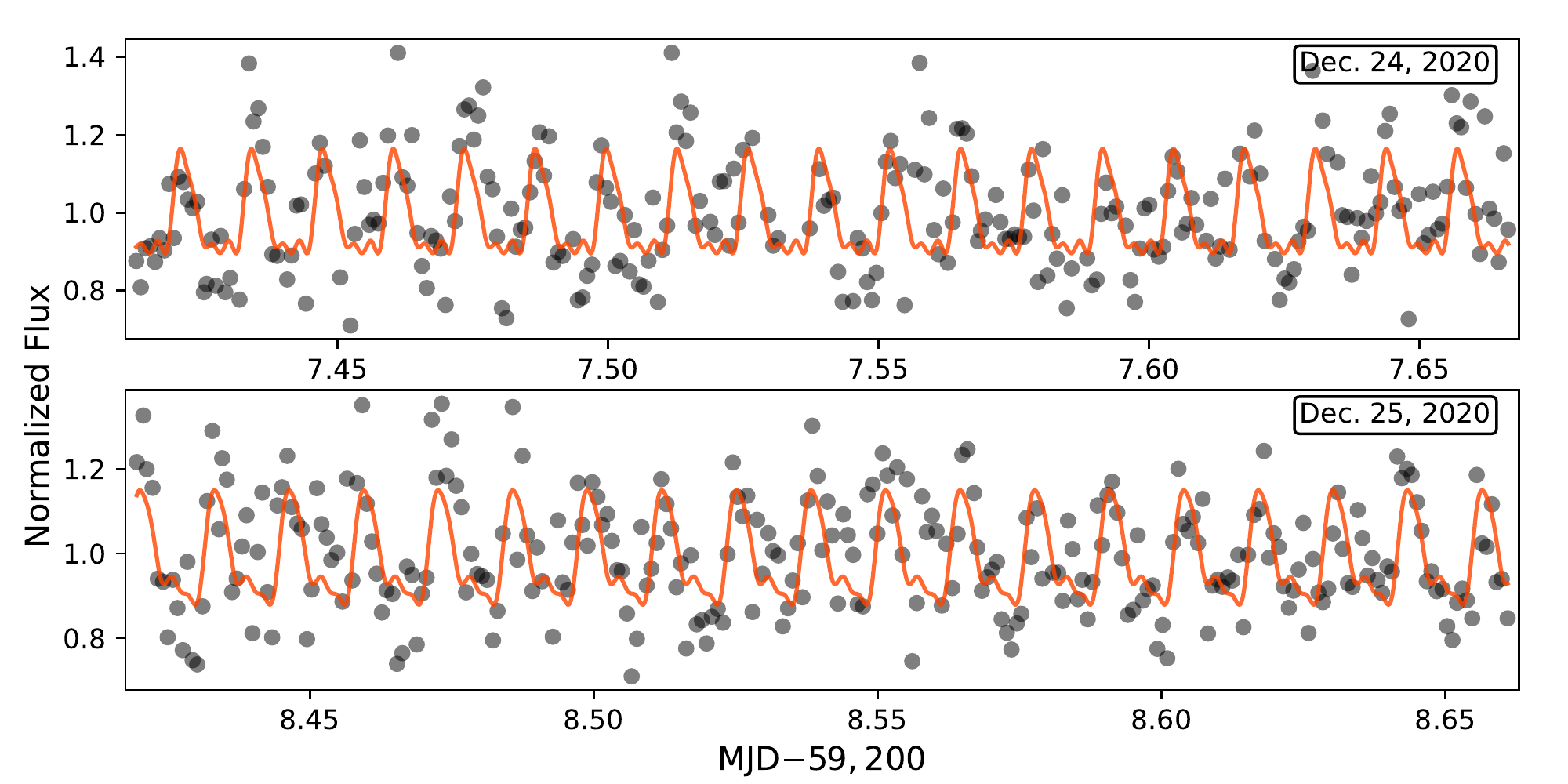}
    \caption{
     TMTS light curves of TMTS-BLAP-1 at $L$ band (white light) on December 24 and 25 , 2020 (UT). The red solid lines represent the best-fitting models of Fourier series truncated at fourth harmonic.
     } 
    \label{fig:tmts_lightcurves}
\end{figure*}
\newpage

\begin{figure*}[]
\centering
    \includegraphics[width=0.98\textwidth]{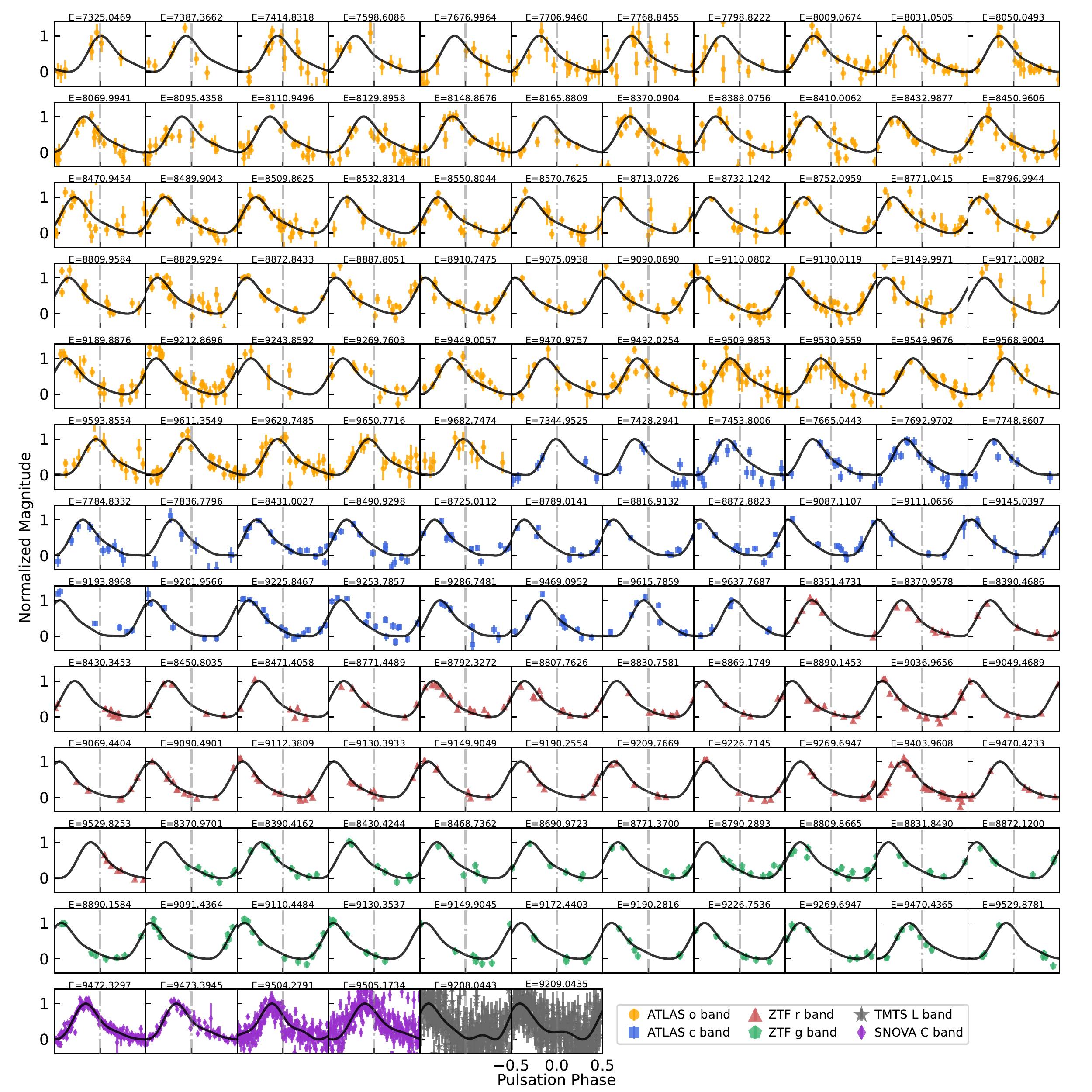}
    \caption{
     Phase-folded light curves for every subset of ATLAS, ZTF, TMTS, and SNOVA data. Every subset of ATLAS and ZTF data covers up to 20~days, while each subset of TMTS and SNOVA data covers only one night.
     The observed time of maximum light ($T_{\rm max}^O$ -- 2,450,000) for every subset is shown above the plots. Since the phases here were all calculated using the ephemeris of Eq.~\ref{Eq:ephemeris}, the pulsation phases $\phi=0$ (the vertical dot-dashed lines) here correspond to the calculated times of maximum light, namely $T_{\rm max}^C$.
     } 
    \label{fig:folded_lc_sections}
\end{figure*}
\newpage

\begin{figure*}[]
\centering
    \includegraphics[width=0.98\textwidth]{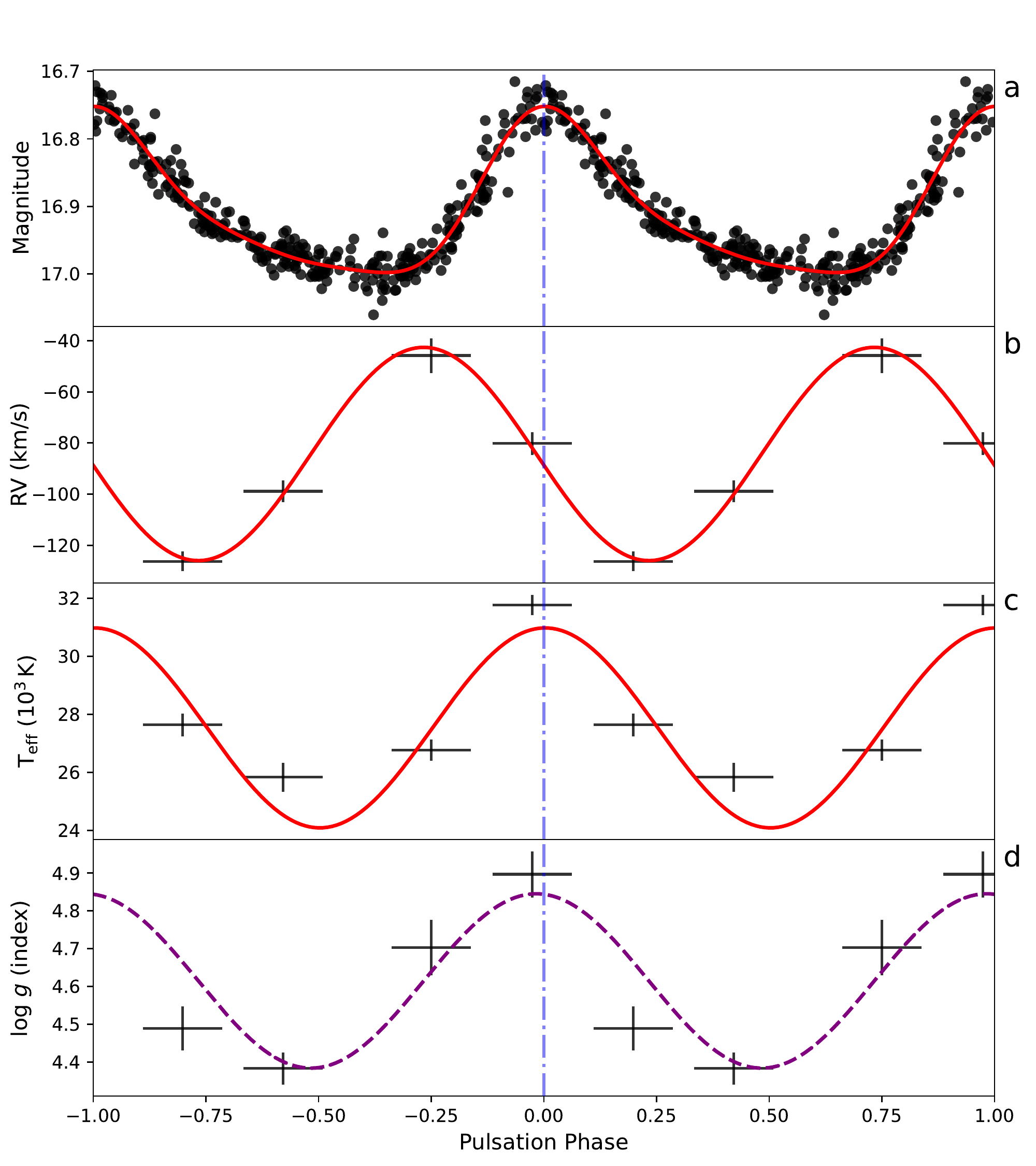}
    \caption{
     Folded light curve and surface parameters against pulsation phase. $a$, corrected ZTF r folded light curve with a best-fitting 3-harmonic Fourier model (red solid line); $b$,$c$,$d$, radial velocity (RV), effective temperature ($T_{\rm eff}$) and surface gravity ($\log\,g$) against pulsation phase. The red solid curves are the best-fitting sinusoidal curves, and the purple dashed line in panel $d$ represents the prediction from the time-derivative of the best-fitting model of radial velocity \cite{Kupfer+2019+high-g_BLAPs}.
     } 
    \label{fig:surface_pars}
\end{figure*}
\newpage

\begin{figure*}[]
\centering
    \includegraphics[width=0.8\textwidth]{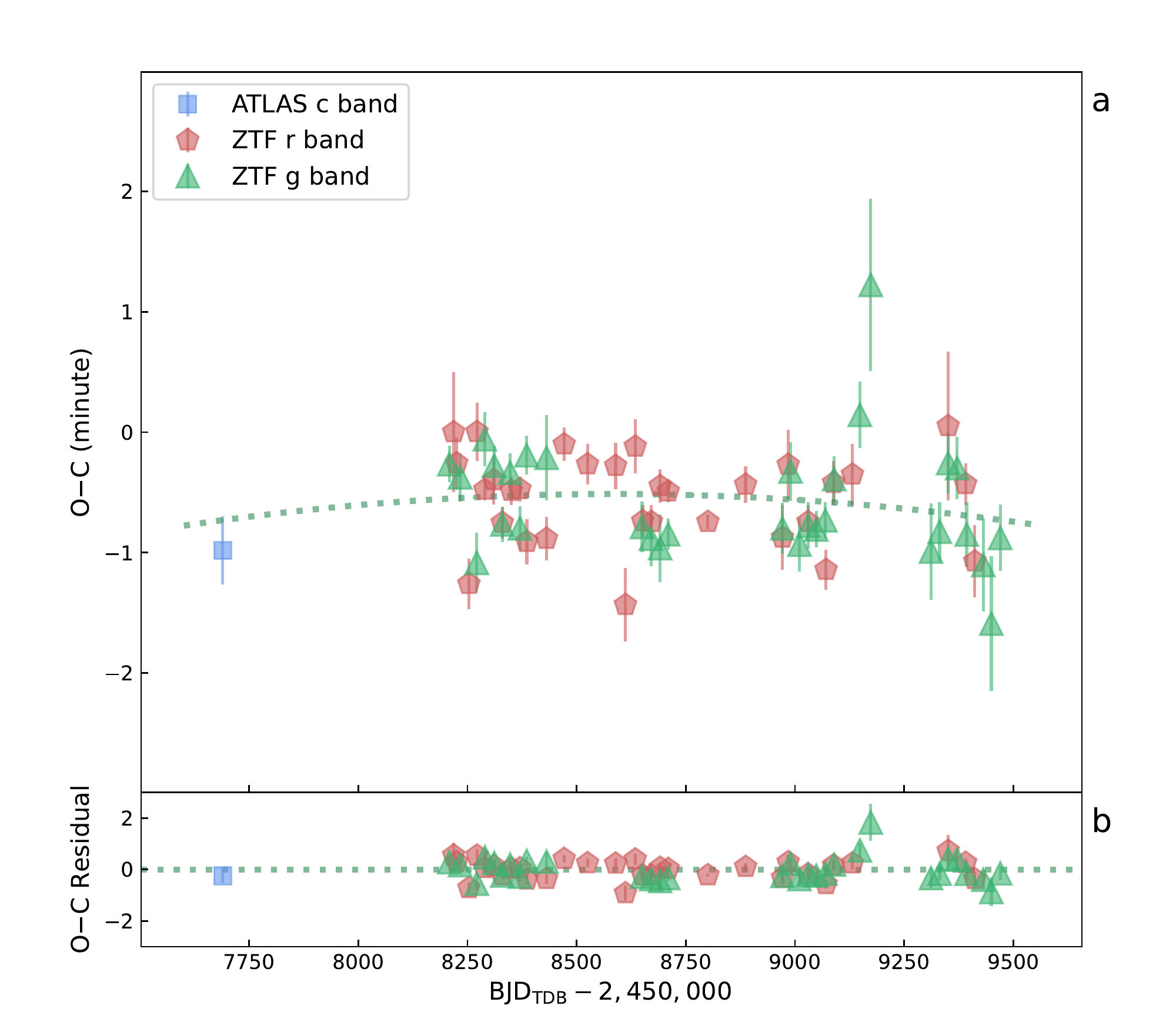}
    \caption{
    $O-C$ diagram for the pulsation period of ZGP-BLAP-09.
The observed time of maximum light ($T_{\rm max}^O$) was obtained from the 20~day subsets of ATLAS and ZTF.
The $O-C$ values were calculated following the the ephemeris
$T_{\rm max}^C(E)={\rm BJD_{TDB}}2,458,218.5012+0.0161558353\times E $.
Because ZGP-BLAP-09 lacks similar cyclic behavior in the diagram, the $O-C$ variability is modeled only by assuming the linear period change.
     } 
    \label{fig:O-C_ztfblap}
\end{figure*}
\newpage

 \begin{figure*}[]
\centering
    \includegraphics[width=0.98\textwidth]{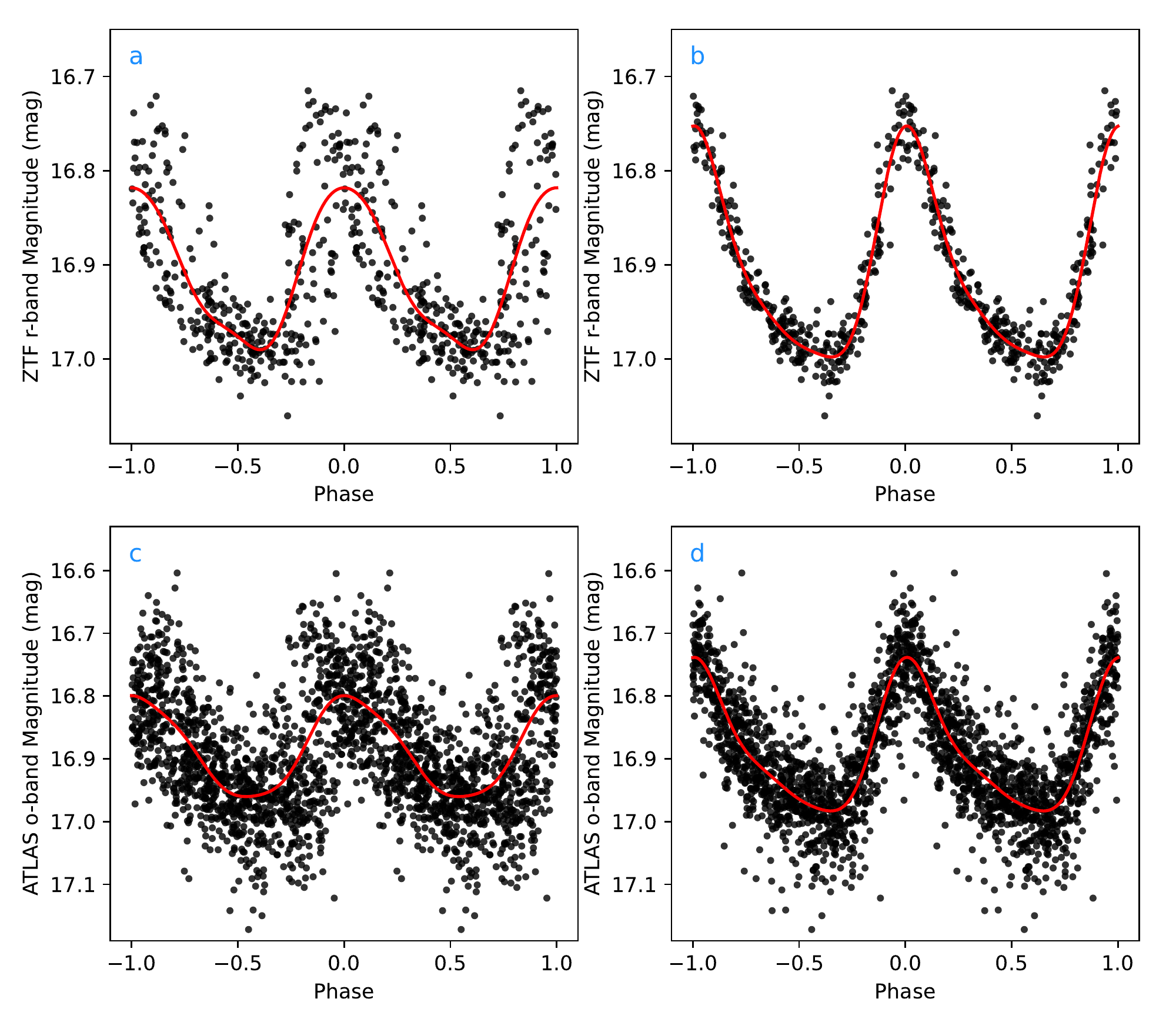}
    \caption{
Phase-folded light curves of TMTS-BLAP-1. The light curves were obtained from ZTF $r$-band (panels $a,b$) and ATLAS $o$-band (panels $c,d$) observations.
     ($a,c$) The light curves are folded using a constant period inferred from the Lomb–Scargle periodogram.
     ($b,d$) The light curves are folded using the new ephemeris derived from the $O-C$ diagram.
     The red solid lines represent the best-fitting 3-harmonic Fourier models.
     } 
    \label{fig:folded_lc}
\end{figure*}
\newpage

\begin{figure*}[]
\centering
    \includegraphics[width=\textwidth]{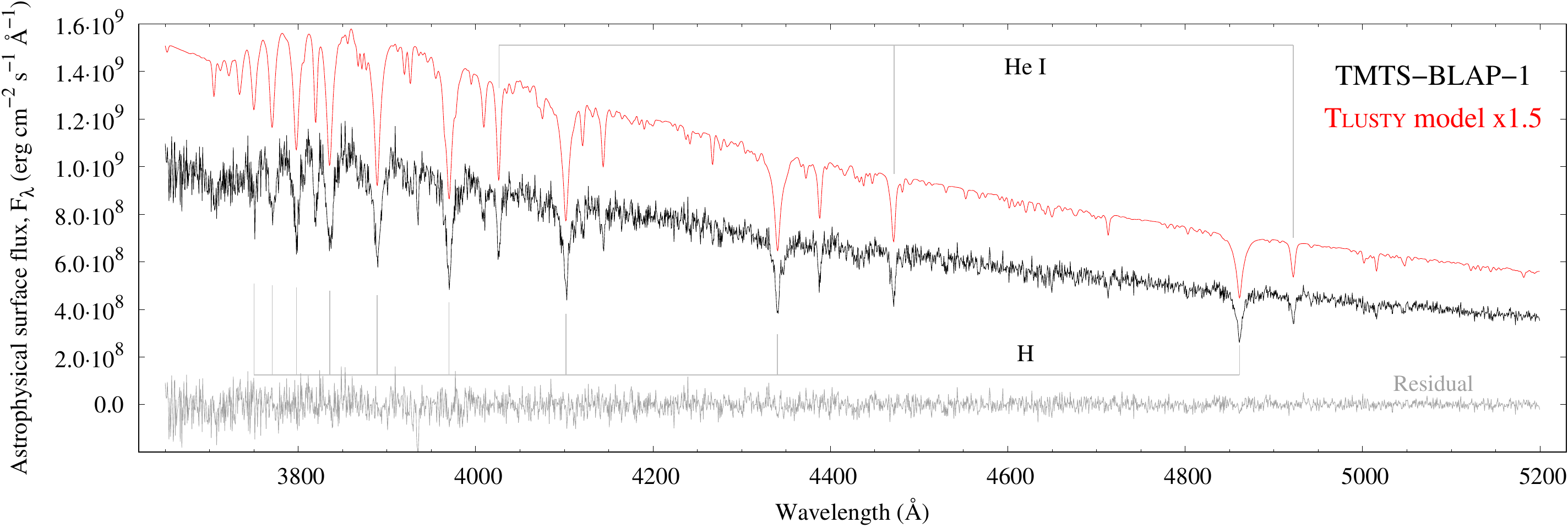}
    \caption{
    Keck-I LRIS spectrum of TMTS-BLAP-1 with its best-fitting {\sc Tlusty/XTgrid} model. 
    The model (red line) is shifted by 50\% for clarity. The grey line represents the residual between observation and model.
     } 
    \label{fig:nlte_fitting}
\end{figure*}

\begin{figure*}[]
\centering
    \includegraphics[width=0.98\textwidth]{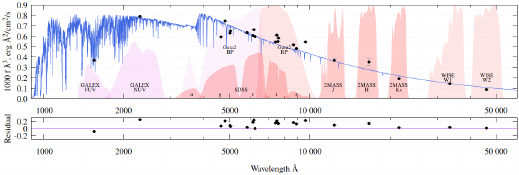}
    \caption{
     Broad-band photometric data from VizieR compared to the SED of TMTS-BLAP-1. 
     %The SED is normalized to the photometry in the {\it Gaia} DR2 $G$ band.
     For better visualization of the UV and infrared part, the flux $f$ was multiplied by $\lambda ^3$.
     The bottom panel shows the residuals after subtracting the synthetic photometry from the observed data.
     }
    \label{fig:SED_fitting}
\end{figure*}

\newpage

\begin{table}[h]
\centering
\caption{ Surface parameters from the four Keck-I LRIS spectra. \label{tab:params}}
\begin{tabular}{lcccc}
\hline
Spectrum:   &  sp0  & sp1   & sp2   & sp3   \\
Pulsation phase:    & $-0.338$ to $-0.162$ & $-0.114$ to $0.062$ & 0.110 to 0.286 & 0.334 to 0.510 \\
\hline
\multicolumn{1}{l|}{Parameter} &&&& \\
\multicolumn{1}{l|}{$T_{\rm eff}$ (K)}           &$26,770\pm370$   &$31,780\pm350$   & $27,640\pm390$   & $25,840\pm500$\\
\multicolumn{1}{l|}{$\log{g}$ (cm\,s$^{-2}$)}    &$4.703\pm0.074$ &$4.897\pm0.061$ & $4.489\pm0.058$ & $4.383\pm0.042$\\
\multicolumn{1}{l|}{$\log{(n{\rm He}/n{\rm H})}$}&$-0.615\pm0.040$&$-0.659\pm0.047$& $-0.620\pm0.055$& $-0.571\pm0.063$\\
\multicolumn{1}{l|}{$\log{(n{\rm C}/n{\rm H})}$ }&$-2.604\pm0.077$&$-2.568\pm0.074$& $-2.624\pm0.071$& $-2.754\pm0.114$\\
\multicolumn{1}{l|}{$\log{(n{\rm N}/n{\rm H})}$ }&$-2.737\pm0.061$&$-3.015\pm0.189$& $-2.633\pm0.097$& $-2.842\pm0.096$\\
\multicolumn{1}{l|}{$\log{(n{\rm O}/n{\rm H})}$ }&$-3.076\pm0.106$&$-2.908\pm0.070$& $-3.004\pm0.147$& $-2.958\pm0.135$\\
\multicolumn{1}{l|}{$\log{(n{\rm Si}/n{\rm H})}$}&$-4.433\pm0.359$&$-4.035\pm0.231$& $-4.500\pm0.317$& $-4.030\pm0.215$\\
\multicolumn{1}{l|}{$\log{(n{\rm Fe}/n{\rm H})}$}&$-4.080\pm0.306$&$-3.918\pm0.453$& $-4.333\pm0.369$& $-3.774\pm0.374$\\
\multicolumn{1}{l|}{RV (km\,s$^{-1}$)}           &$-45.8\pm6.8$   &$-80.2\pm4.5$   & $-126.3\pm3.9$  & $-98.9\pm4.2$\\
\multicolumn{1}{l|}{{\it Gaia3} $(B_P-R_P)_0$}& $-0.42$ & $-0.47$ & $-0.42$ & $-0.41$ \\
\hline
\end{tabular}
\end{table}
\newpage

\begin{table}[ht]
%\setstretch{1.1}
\centering
\caption{Fluxes from the broad-band photometric data and the corresponding synthetic fluxes. All measurements were extracted from the VizieR SED service. The theoretical and observed fluxes were matched in the {\it Gaia} DR2 $G$ band. \label{tab:sed} }
\begin{tabular}{lrrr}
\hline
Filter& $\lambda_{\rm eff}$ & $F_\lambda$ (obs) & $F_\lambda$ (syn)\\
      & (\AA)               & (erg/cm$^2$/s/\AA)  & (erg/cm$^2$/s/\AA) \\
\hline
GALEX:FUV  &1549 &9.96e-14 &1.55e-13\\
GALEX:NUV  &2304 &6.49e-14 &4.96e-14\\
SDSS:g     &4640 &5.95e-15 &5.43e-15\\
PS1:g      &4810 &6.71e-15 &4.93e-15\\
Gaia3:BP   &5035 &5.11e-15 &4.83e-15\\
Gaia2:BP   &5017 &5.00e-15 &4.48e-15\\
Gaia3:G    &5822 &3.22e-15 &3.17e-15\\
SDSS:r     &6122 &2.65e-15 &1.89e-15\\
PS1:r      &6156 &2.85e-15 &1.89e-15\\
{\bf Gaia2:G} &{\bf 6230} &{\bf 2.48e-15} & {\bf 2.48e-15} \\
SDSS:i     &7439 &1.32e-15 &9.92e-16\\
PS1:i      &7503 &1.44e-15 &9.74e-16\\
Gaia2:RP   &7593 &1.26e-15 &9.10e-16\\
Gaia3:RP   &7619 &1.31e-15 &9.64e-16\\
PS1:z      &8668 &7.98e-16 &5.48e-16\\
SDSS:z     &8897 &6.84e-16 &4.92e-16\\
PS1:y      &9613 &6.15e-16 &3.71e-16\\
2MASS:J   &12350 &1.97e-16 &1.49e-16\\
2MASS:H   &16620 &7.68e-17 &4.62e-17\\
2MASS:Ks  &21590 &1.91e-17 &1.70e-17\\
WISE:W1   &33526 &3.85e-18 &3.16e-18\\
WISE:W2   &46028 &9.05e-19 &8.90e-19\\
\hline
\end{tabular}
\end{table}

\newpage

\begin{longtable}{lrrrrrr}
\caption{Times of maximum light and $O-C$ for TMTS-BLAP-1. The $T_{\rm max}$ are given as ${\rm BJD_{TDB}}-$2,450,000.   \label{tab:Tmax}}\\
\hline\hline
\multicolumn{1}{c}{$T_{\rm max}$} & \multicolumn{1}{c}{Error} & \multicolumn{1}{c}{$E$} & \multicolumn{1}{c}{$O-C$} & \multicolumn{1}{c}{$(O-C)_{\rm res}$} & \multicolumn{1}{c}{Instrument/Band}  \\
& \multicolumn{1}{c}{(minute)} &  & \multicolumn{1}{c}{(minute)} & \multicolumn{1}{c}{(minute)} & \\
\hline
\tiny
\endfirsthead
\caption{continued.}\\
\hline\hline
\multicolumn{1}{c}{$T_{\rm max}$} & \multicolumn{1}{c}{Error} & \multicolumn{1}{c}{$E$} & \multicolumn{1}{c}{$O-C$} & \multicolumn{1}{c}{$(O-C)_{\rm res}$}& \multicolumn{1}{c}{Instrument/Band} \\
& \multicolumn{1}{c}{(minute)} &  & \multicolumn{1}{c}{(minute)} & \multicolumn{1}{c}{(minute)} & \\
\hline
\endhead
\hline
\endfoot

7325.04694&1.05&0&0.00&0.03&ATLAS/o\\
7344.95249&0.39&1514&-0.14&0.29&ATLAS/c\\
7387.36620&0.34&4740&-1.31&-0.05&ATLAS/o\\
7414.83181&0.32&6829&-1.26&0.53&ATLAS/o\\
7428.29406&0.48&7853&-2.72&-0.69&ATLAS/c\\
7453.80062&0.29&9793&-2.73&-0.23&ATLAS/c\\
7598.60863&0.88&20807&-4.06&-0.08&ATLAS/o\\
7665.04426&0.37&25860&-3.73&0.03&ATLAS/c\\
7676.99644&0.71&26769&-2.43&1.28&ATLAS/o\\
7692.97024&0.22&27984&-3.40&0.23&ATLAS/c\\
7706.94600&0.56&29047&-3.78&-0.22&ATLAS/o\\
7748.86072&0.63&32235&-4.06&-0.68&ATLAS/c\\
7768.84552&0.47&33755&-3.67&-0.36&ATLAS/o\\
7784.83322&0.32&34971&-3.56&-0.30&ATLAS/c\\
7798.82219&0.54&36035&-3.85&-0.62&ATLAS/o\\
7836.77959&0.51&38922&-3.91&-0.75&ATLAS/c\\
8009.06736&0.25&52026&-3.76&-0.53&ATLAS/o\\
8031.05049&0.19&53698&-3.55&-0.27&ATLAS/o\\
8050.04929&0.22&55143&-3.03&0.28&ATLAS/o\\
8069.99409&0.21&56660&-3.44&-0.07&ATLAS/o\\
8095.43581&0.27&58595&-2.16&1.28&ATLAS/o\\
8110.94964&0.18&59775&-2.84&0.64&ATLAS/o\\
8129.89584&0.15&61216&-2.35&1.20&ATLAS/o\\
8148.86765&0.24&62659&-2.85&0.77&ATLAS/o\\
8165.88095&0.25&63953&-2.62&1.05&ATLAS/o\\
8351.47308&0.15&78069&-4.09&0.39&ZTF/r\\
8370.09036&0.19&79485&-3.91&0.66&ATLAS/o\\
8370.95778&0.28&79551&-4.39&0.18&ZTF/r\\
8370.97010&0.43&79552&-5.58&-1.01&ZTF/g\\
8388.07561&0.22&80853&-5.10&-0.44&ATLAS/o\\
8390.41620&0.26&81031&-4.68&-0.00&ZTF/g\\
8390.46861&0.24&81035&-4.93&-0.26&ZTF/r\\
8410.00616&0.17&82521&-4.87&-0.10&ATLAS/o\\
8430.34527&0.20&84068&-5.45&-0.57&ZTF/r\\
8430.42436&0.58&84074&-5.16&-0.28&ZTF/g\\
8431.00266&0.15&84118&-5.45&-0.57&ATLAS/c\\
8432.98770&0.22&84269&-5.83&-0.93&ATLAS/o\\
8450.80347&0.21&85624&-4.95&0.04&ZTF/r\\
8450.96064&0.23&85636&-5.81&-0.82&ATLAS/o\\
8468.73618&0.54&86988&-6.05&-0.96&ZTF/g\\
8470.94541&0.19&87156&-5.47&-0.37&ATLAS/o\\
8471.40581&0.23&87191&-5.13&-0.02&ZTF/r\\
8489.90429&0.16&88598&-5.64&-0.43&ATLAS/o\\
8490.92977&0.21&88676&-5.70&-0.48&ATLAS/c\\
8509.86246&0.20&90116&-5.74&-0.42&ATLAS/o\\
8532.83143&0.36&91863&-5.86&-0.41&ATLAS/o\\
8550.80439&0.19&93230&-5.81&-0.25&ATLAS/o\\
8570.76250&0.30&94748&-5.98&-0.31&ATLAS/o\\
8690.97232&0.39&103891&-5.61&0.80&ZTF/g\\
8713.07259&0.87&105572&-7.11&-0.56&ATLAS/o\\
8725.01122&0.16&106480&-6.37&0.26&ATLAS/c\\
8732.12423&0.24&107021&-6.24&0.43&ATLAS/o\\
8752.09591&0.24&108540&-5.80&1.00&ATLAS/o\\
8771.04150&0.20&109981&-6.19&0.74&ATLAS/o\\
8771.37003&0.37&110006&-6.43&0.49&ZTF/g\\
8771.44890&0.23&110012&-6.45&0.47&ZTF/r\\
8789.01406&0.19&111348&-6.71&0.32&ATLAS/c\\
8790.28927&0.38&111445&-6.89&0.16&ZTF/g\\
8792.32719&0.13&111600&-6.85&0.21&ZTF/r\\
8796.99442&0.23&111955&-7.15&-0.06&ATLAS/o\\
8807.76263&0.15&112774&-6.82&0.34&ZTF/r\\
8809.86646&0.21&112934&-6.54&0.63&ZTF/g\\
8809.95841&0.30&112941&-6.67&0.51&ATLAS/o\\
8816.91323&0.26&113470&-7.13&0.09&ATLAS/c\\
8829.92944&0.18&114460&-7.16&0.14&ATLAS/o\\
8830.75810&0.22&114523&-6.65&0.65&ZTF/r\\
8831.84902&0.30&114606&-7.15&0.17&ZTF/g\\
8869.17493&0.18&117445&-7.79&-0.24&ZTF/r\\
8872.12002&0.25&117669&-7.79&-0.22&ZTF/g\\
8872.84327&0.22&117724&-7.61&-0.03&ATLAS/o\\
8872.88235&0.20&117727&-8.14&-0.56&ATLAS/c\\
8887.80506&0.23&118862&-8.06&-0.38&ATLAS/o\\
8890.14529&0.23&119040&-8.15&-0.46&ZTF/r\\
8890.15840&0.33&119041&-8.20&-0.51&ZTF/g\\
8910.74752&0.20&120607&-8.49&-0.67&ATLAS/o\\
9036.96564&0.11&130207&-8.42&0.05&ZTF/r\\
9049.46893&0.21&131158&-8.68&-0.19&ZTF/r\\
9069.44037&0.18&132677&-8.60&-0.07&ZTF/r\\
9075.09376&0.30&133107&-8.78&-0.25&ATLAS/o\\
9087.11073&0.23&134021&-8.84&-0.31&ATLAS/c\\
9090.06899&0.19&134246&-8.82&-0.29&ATLAS/o\\
9090.49006&0.17&134278&-8.31&0.22&ZTF/r\\
9091.43636&0.17&134350&-8.80&-0.28&ZTF/g\\
9110.08023&0.19&135768&-8.21&0.28&ATLAS/o\\
9110.44841&0.19&135796&-8.14&0.34&ZTF/g\\
9111.06558&0.37&135843&-9.26&-0.78&ATLAS/c\\
9112.38088&0.15&135943&-8.50&-0.02&ZTF/r\\
9130.01193&0.18&137284&-8.55&-0.17&ATLAS/o\\
9130.35368&0.30&137310&-8.67&-0.29&ZTF/g\\
9130.39327&0.16&137313&-8.46&-0.09&ZTF/r\\
9145.03973&0.23&138427&-8.61&-0.35&ATLAS/c\\
9149.90446&0.33&138797&-8.49&-0.28&ZTF/g\\
9149.90490&0.23&138797&-7.87&0.35&ZTF/r\\
9149.99711&0.28&138804&-7.61&0.60&ATLAS/o\\
9171.00817&0.56&140402&-6.16&1.81&ATLAS/o\\
9172.44032&0.31&140511&-7.52&0.42&ZTF/g\\
9189.88759&0.26&141838&-7.16&0.53&ATLAS/o\\
9190.25542&0.14&141866&-7.60&0.09&ZTF/r\\
9190.28160&0.31&141868&-7.76&-0.08&ZTF/g\\
9193.89683&0.25&142143&-8.32&-0.70&ATLAS/c\\
9201.95658&0.36&142756&-8.04&-0.55&ATLAS/c\\
9208.04432&0.22&143219&-7.54&-0.16&TMTS/L\\
9209.04347&0.22&143295&-7.64&-0.28&TMTS/L\\
9209.76692&0.39&143350&-7.18&0.17&ZTF/r\\
9212.86956&0.16&143586&-7.49&-0.19&ATLAS/o\\
9225.84669&0.14&144573&-7.01&0.06&ATLAS/c\\
9226.71450&0.24&144639&-6.93&0.12&ZTF/r\\
9226.75361&0.30&144642&-7.40&-0.35&ZTF/g\\
9243.85921&0.39&145943&-6.79&-0.06&ATLAS/o\\
9253.78569&0.17&146698&-6.87&-0.32&ATLAS/c\\
9269.69466&0.21&147908&-6.53&-0.28&ZTF/g\\
9269.69472&0.15&147908&-6.44&-0.20&ZTF/r\\
9269.76031&0.25&147913&-6.65&-0.41&ATLAS/o\\
9286.74811&0.53&149205&-5.28&0.65&ATLAS/c\\
9403.96080&0.08&158120&-4.11&-0.19&ZTF/r\\
9449.00565&0.19&161546&-2.99&0.26&ATLAS/o\\
9469.09522&0.24&163074&-3.19&-0.22&ATLAS/c\\
9470.42333&0.25&163175&-2.91&0.04&ZTF/r\\
9470.43651&0.22&163176&-2.87&0.08&ZTF/g\\
9470.97572&0.20&163217&-2.66&0.29&ATLAS/o\\
9472.32968&0.06&163320&-3.02&-0.10&SNOVA/C\\
9473.39455&0.07&163401&-3.16&-0.25&SNOVA/C\\
9492.02538&0.17&164818&-2.41&0.24&ATLAS/o\\
9504.27912&0.06&165750&-2.31&0.18&SNOVA/C\\
9505.17339&0.12&165818&-1.99&0.49&SNOVA/C\\
9509.98534&0.26&166184&-2.15&0.27&ATLAS/o\\
9529.82526&0.24&167693&-2.13&0.03&ZTF/r\\
9529.87814&0.32&167697&-1.71&0.45&ZTF/g\\
9530.95586&0.25&167779&-2.28&-0.13&ATLAS/o\\
9549.96763&0.22&169225&-2.03&-0.13&ATLAS/o\\
9568.90035&0.26&170665&-2.00&-0.34&ATLAS/o\\
9593.85543&0.33&172563&-0.98&0.39&ATLAS/o\\
9611.35491&0.22&173894&-1.17&-0.01&ATLAS/o\\
9615.78587&0.21&174231&-0.90&0.21&ATLAS/c\\
9629.74850&0.16&175293&-1.25&-0.30&ATLAS/o\\
9637.76869&0.65&175903&-1.12&-0.27&ATLAS/c\\
9650.77163&0.23&176892&-1.34&-0.63&ATLAS/o\\
9682.74736&1.00&179324&-0.65&-0.29&ATLAS/o\\
\noalign{\smallskip}\hline 
\end{longtable}

\newpage

\end{document}